\documentclass[superscriptaddress,groupedaddress,nofootnoteinbib,11pt]{article}
\pdfoutput=1 
\usepackage{jheppub}

\usepackage{amsmath, amsfonts, amsthm, amssymb, graphicx, xcolor, hyperref, mathrsfs}
\usepackage{caption}
\usepackage{subcaption}
\usepackage{float}

\usepackage[utf8]{inputenc}

\allowdisplaybreaks[3]

\def\({\left(}
\def\){\right)}
\def\[{\left[}
\def\]{\right]}
\def\d{\mathrm{d}}

\newcommand{\p} {\partial}
\newcommand{\f}[2]{\frac{#1}{#2}}
\def \bal#1\eal  {\begin{align} #1 \end{align}}
\newcommand{\eref}[1]{{Eq.~(\ref{#1})}}
\newcommand{\be} {\begin{equation}}
\newcommand{\ee} {\end{equation}}
\newcommand{\bc}{\begin{center}}
	\newcommand{\ec}{\end{center}}
\newcommand{\bim} {\begin{itemize}[noitemsep]}
	
	\newcommand{\eim} {\end{itemize}}
\newcommand{\lt}{\left}
\newcommand{\rt}{\right}

\newcommand{\ie}{{\it i.e.}, }

\newcommand{\pd} {\partial}
\newcommand{\mc} {\mathcal}

      \newcommand{\bfx} {{\bf x}}

\newcommand{\ai}{{\alpha}}

\newcommand{\ri}{{\rho}}
\newcommand{\si}{{\sigma}}

\newcommand{\oi}{\omega}

\newcommand{\cphi}{\varphi}
\newcommand{\cepi}{\varepsilon}
 
\graphicspath{ {./images/}{./CSQ/}{./gaugedCSQ/}{./hsyresult/} }

\title{Charge-Swapping Q-balls in a Logarithmic Potential and Affleck-Dine condensate fragmentation}

\author[a]{Si-Yuan Hou,}
\author[b]{Paul M. Saffin,}
\author[c,d]{Qi-Xin Xie,}
\author[c,d]{and Shuang-Yong Zhou}

\affiliation[a]{College of Astronautics, Nanjing University of Aeronautics and Astronautics,\\ Nanjing, Jiangsu 210016, China}
\affiliation[b]{School of Physics and Astronomy, University Park, University of Nottingham,\\ Nottingham NG7 2RD, United Kingdom}
\affiliation[c]{Interdisciplinary Center for Theoretical Study, University of Science and Technology of China,\\ Hefei, Anhui 230026, China}
\affiliation[d]{Peng Huanwu Center for Fundamental Theory, Hefei, Anhui 230026, China}

\emailAdd{housiyuan2001@outlook.com}
\emailAdd{paul.saffin@nottingham.ac.uk}
\emailAdd{xqx2018@mail.ustc.edu.cn}
\emailAdd{zhoushy@ustc.edu.cn}
\preprint{\small USTC-ICTS/PCFT-22-04}

\date{\today}

\abstract{
We study charge-swapping Q-balls, a kind of composite Q-ball where positive and negative charges co-exist and swap with time, in models with a logarithmic potential that arises naturally in supersymmetric extensions of the Standard Model. We show that charge-swapping Q-balls can be copiously generated in the Affleck-Dine fragmentation process in the early universe. We find that the charge-swapping Q-balls with the logarithmic potential are extremely stable. By performing long time, parallelized lattice simulations with absorbing boundary conditions, we find that the lifetimes of such objects with low multipoles are at least $4.6 \times 10^5/m$ in 3+1D and $2.5 \times 10^7/m$ in 2+1D, where $m$ is the mass scale of the scalar field. We also chart the attractor basin of the initial conditions to form these charge-swapping Q-balls.
}

\begin{document}
	\maketitle
	\flushbottom

	\section{Introduction}
	\label{sec:introduction}

Q-balls are a type of non-topological soliton \cite{Rosen:1968mfz,Friedberg:1976me,Coleman:1985ki}, which are nonlinear field configurations confined in a finite space region and are stationary, rather than static, in time. The stability of Q-balls is usually guaranteed by some Noether charges, rather than topological charges in the case of topological solitons \cite{Manton:2004tk} whose configurations can usually be static. Q-balls can arise in a large number of field theories, the conditions for them to exist being quite general. For instance, Q-balls can exist in a self-interacting complex scalar field theory whose potential grows slower than the quadratic mass term away from the minimum. There are many flat directions in the potential of the supersymmetric extensions of the Standard Model, and these flat directions can support Q-balls \cite{Kusenko:1997zq, Laine:1998rg,Enqvist:2000gq}. Indeed, in these supersymmetric models, Q-balls can naturally form in the early universe as a by-product of Affleck-Dine baryogenesis \cite{Affleck:1984fy,Enqvist:1997si,Enqvist:1998en,Enqvist:1999mv,Kasuya:2000wx,Multamaki:2002hv,Harigaya:2014tla}, as they are energetically preferred states in these potentials. Q-balls copiously generated in the early universe can survive to the present epoch and act as a dark matter candidate \cite{Kusenko:1997si,Enqvist:1998xd,Banerjee:2000mb,Kusenko:2001vu,Roszkowski:2006kw,Shoemaker:2009kg,Kasuya:2011ix,Kasuya:2012mh,Kawasaki:2019ywz}. They can also be macroscopic compact objects with strong self-gravity, in which case they usually go under the name of boson stars \cite{Visinelli:2021uve}.

While most of the literature has focused on the simplest Q-ball solutions where its spatial profile is spherically symmetric, Ref \cite{Copeland:2014qra} revealed the surprising existence of a tower of composite Q-balls, which are quasi-spherical in energy density but multipolar in charge density and whose constituent positive and negative charges swap in time. They were dubbed charge-swapping Q-balls (CSQs), and are long-lived, meta-stable states, which exist in theories when the simple spherical Q-ball solution exists. A CSQ can be easily prepared by placing positive and negative charge lumps tightly together, which quickly relaxes and evolves to the quasi-stable CSQ configuration, as the CSQ is a meta-stable attractor solution. As shown in the case of a U(1) scalar theory with a sextic polynomial potential, after the initial relaxation, we end up with a CSQ that lasts about $10^4$ oscillations in 2D and about $10^3$ oscillations in 3D for a sextic dimensionless coupling of $1/2$ (the lifetime increases exponentially for larger sextic couplings), during which stage the charge-swapping frequency of a CSQ remains mostly constant \cite{Xie:2021glp}. Afterwards, the quasi-stable CSQ suddenly decays into an oscillon \cite{Bogolyubsky:1976yu,Copeland:1995fq}, during which stage the remaining small charges decay exponentially and the resulting configuration mostly oscillates along one direction in the U(1) field space. See \cite{Mai:2012cx, Abel:2015tca, Bazeia:2016xrf, Kovtun:2018jae, Loiko:2018mhb, Panin:2018uoy, Hasegawa:2019bbo, Loginov:2020xoj, VanDissel:2020umg, Almumin:2021gax, Klimas:2022ghu, Lennon:2021fde} for other forms of excited Q-balls and Q-ball-like configurations.
	
In this paper, we shall investigate formation, dynamics and longevity of CSQs in scalar theories with a logarithmic potential (see \eref{eq:log potential}). The logarithmic potential is a typical quantum effective potential that naturally arises in supersymmetric extensions of the Standard Model such as in the gravity-mediated supersymmetric breaking scenario \cite{Enqvist:1997si,Enqvist:1998en,Enqvist:1999mv,Multamaki:1999an,Enqvist:2000gq,Kasuya:2000wx} or any scenario where the quantum corrected mass term decreases with the energy scale.  The attractor nature of CSQs means that they can be generated from random initial conditions in the early universe. As we shall see, if a U(1) scalar field forms a VEV, say an Affleck-Dine condensate \cite{Affleck:1984fy}, in the early universe, after the Hubble scale drops below the mass of the scalar, the condensate can fragment into positive and negative charge lumps, which creates the perfect breeding ground for CSQs to form. Different from the sextic polynomial potential studied in \cite{Xie:2021glp}, which is the simplest potential that supports Q-balls, the logarithmic potential is fascinating because it is ultra-soft, distinct from all other potentials. 

To accurately determine the lifetimes of logarithmic CSQs, we focus on CSQs with low multipoles, which can be prepared in a more controlled form, by superimposing opposite charge lumps such that their nonlinear cores overlap. The usual choice of periodic boundary conditions is unsuitable for this purpose because during the initial relaxation and in the CSQ stage perturbations, or radiation emitted by the CSQs, can introduce artificial instabilities into the system. Instead, we shall implement the effective Higdon absorbing boundary conditions \cite{higdon1994radiation, higdon1986absorbing} in our lattice simulations. A distinct feature due to the ultra-softness of the logarithmic potential is that the relaxation process emits far less radiation than the case of the sextic potential. Once formed, the logarithmic CSQs are extremely stable. In fact, we have yet to capture their decay in our long-time parallelized simulations: in 3+1D simulations our longest run time is $4.6 \times 10^5/m$ or about $1.1\times 10^5$ CPU hours, and in 2+1D the longest run time is $2.5 \times 10^7/m$ or about $2\times10^4$ CPU hours, where $m$ is the mass scale in the quadratic term. CSQs with unequal opposite charges are also shown to be very stable, at least for the relatively simple ones. We also investigate the conditions for forming logarithmic CSQs, performing a comprehensive survey of the parameter space for formation with frequencies, separations and velocities of the initial lumps. We see that logarithmic CSQs are relatively easy to form, even when colliding opposite charge lumps with relatively large impact parameters and/or initial velocities, and there are a few distinct outcomes of CSQ formation for non-head-on collisions.

The paper is organized as follows. In Section \ref{sec:model}, we introduce the model we will study in this paper, a U(1) scalar field with a logarithmic potential. This is a typical effective potential where the mass term receives logarithmic quantum corrections and which, for example, can come from supersymmetric extensions of the Standard Model. In Section \ref{sec:CSQfromAD}, we show 
that in such a scenario an Affleck-Dine condensate, a VEV formed in the early universe, would naturally fragment to complex CSQs, rather than single charge Q-balls. The simulations performed in this section are in an expanding universe and with random initial conditions. In Section \ref{sec:stabiCSQs}, we take a simplified approach to prepare and study CSQs with low multipoles. This provides a controlled way to extract the most salient properties of the logarithmic CSQs. We first outline the absorbing boundary conditions and other numerical setups for our lattice simulations in this section, which are needed to determine the long term stability of these CSQs. The simulations performed in this section are thus with a different code built on the LATfield2 framework from scratch. We then investigate the evolution and dynamics of logarithmic CSQs, and explore the attractor basin for forming logarithmic CSQs. More complex CSQs are also explored in this section. We summarize our results in Section \ref{sec:Summary}.

	\section{Model}
	\label{sec:model}
	
	As mentioned, a class of ultra-soft effective potentials naturally arise in the context of gravity-mediated supersymmetric breaking scenarios, which may be phenomenologically relevant in the very early universe. This class of potentials is characterized by a logarithmic interaction term, which arises from including loop effects that incorporate the running of the potential with energy scale. As the fields along the various flat directions have suppressed couplings between themselves and with other light fields, we can restrict to one complex scalar \cite{Affleck:1984fy, Enqvist:1997si}. Specifically, we consider the following model 
	\be 
	\label{eq:Lagrangian}
	\mc{L}_{\cphi} = -(\p_{\mu}\cphi)^*\p^{\mu}\cphi -V(|\cphi|)  ,
	\ee 
	with
	\be 
	\label{eq:log potential}
	V(|\varphi|) = m^2 |\varphi|^2 \lt( 1+ K \ln  \f{|\varphi|^2}{M^2}  \rt) ,
	\ee 
where $\cphi$ is a complex scalar field,  $m$ is the soft-breaking mass at the scale $M$, and $K$ is a coefficient that is usually taken to be $ -\alpha_{s} m_{1 / 2}^{2} / 8 \pi m_{\tilde{l}}^{2}\sim -0.01 \sim -0.1$, with $\alpha_{s}$, $m_{1 / 2}$ and $m_{\tilde{l}}$ being the coupling of the strong interaction, the gaugino mass and the slepton mass respectively \cite{Enqvist:1997si}.  As we shall do shortly in \eref{eq:dimensionless quantities 1}, we will redefine the coordinates and the fields to absorb $m$ and $M$, so our simulations will be valid for all $m$ and $M$. Of course, this potential may arise as a quantum effective potential from some other models with different values for $K$, and we may take the view of being largely agnostic about the origin of this model, as much of the dynamics of the CSQs are insensitive to moderate deviations of the $K$ values we choose. 

The global U(1) symmetry implies that we have a Noether current $j^{\si}$ in the theory
	\be 
	\label{eq:physical current}
	j^{\si} = 2 {\rm Im} ( \cphi^* \pd^{\si}\cphi ) ,
	\ee 
	where $j^{\si}$ is the charge density, and we can define a conserved charge
	\be 
	Q = \int j^0 \d x^3. 
	\ee 
	The equation of motion for $\cphi$ is given by
	\be
	\label{eq:eom full1}
	\pd_{\mu}\pd^{\mu}\cphi = \f{\p V}{\p \cphi^*}  ,
	\ee 
and the energy density of this system is 
	\be 
	\ri = T_{00} = \( \pd_{\ai}\cphi^*\pd^{\ai}\cphi  +V \) +2 \pd_{0}\cphi^*\pd_{0}\cphi . 
	\label{eq:density}
	\ee 
In the simulations and in the plots of the results, we shall use dimensionless quantities by redefining the following quantities for convenience: 
	\be 
	\label{eq:dimensionless quantities 1}
	mx^{\mu} \to x^{\mu} , \qquad \f{\varphi}{M} \to \varphi .
	\ee 
This is equivalent to taking the coordinates in units of $1/m$, the $\varphi$ values in units of $M$, energy densities in units of $m^2M^2$ and charge densities in units of $mM^2$.

An important feature of this potential is that it grows slower than the quadratic mass term, \ie  its potential provides an attractive force between ``particles'' in this model. This kind of potential can support Q-ball solutions \cite{Friedberg:1976me}, as for a given total U(1) charge the localized Q-ball solution is the minimum of the energy functional \cite{Coleman:1985ki}. That is, in this potential, ``particles'' prefer to stick together to form a spherical ball configuration rather than dissipate to infinity, due to the attractive force. The simple (spherically symmetric) Q-balls of this model have been studied in \cite{Enqvist:1997si, Enqvist:1998en}, and collisions between the simple Q-balls in this model have also been extensively studied previously \cite{MAKHANKOV1979171,Drohm:1981pc,Belova:1988gg,Axenides:1999hs,Battye:2000qj,Multamaki:2000qb,Kasuya:2000wx,Postma:2001ea,Bowcock:2008dn}.

As shown in \cite{Copeland:2014qra}, the structure of a Q-ball can actually be much richer. While the simple Q-ball solution is spherically symmetric, CSQs are composite Q-balls that are excited and quasi-stable, with their energy densities quasi-spherical and their charge densities multipolar. Indeed, they exist in theories where the simple Q-balls exist, and for each theory there is a tower of CSQs with different multipoles. The most significant feature of a CSQ is that its constituent positive and negative charges swap in time, and the swapping frequency, which is smaller than the oscillation frequency of the field, remains mostly constant before the CSQ decays into an oscillon. {\it We can define the CSQ as a state where its charge swapping frequency remains mostly constant for a prolonged period \cite{Xie:2021glp}, at least for low multipolar CSQs.} 

In Section \ref{sec:CSQfromAD}, we will see that complex CSQs are the natural products of the Affleck-Dine fragmentation in the early universe. In Section \ref{sec:stabiCSQs}, we will investigate the properties, dynamics and stability of CSQs with low multipoles.

\section{CSQs from Affleck-Dine fragmentation}
\label{sec:CSQfromAD}

With favorable conditions, the complex scalar field can acquire a VEV or condensate in the early universe, which can make up a significant portion of the energy density in the universe when the condensate starts to oscillate. In Affleck-Dine baryogenesis \cite{Affleck:1984fy}, a baryon number can be generated by additional small A-terms in the potential, which nevertheless become negligible at late times. Q-balls are by-products in this mechanism when the Affleck-Dine condensate fragments. The fragmentation of the Affleck-Dine condensate is a process where initially the condensate oscillates periodically and the (quantum) random perturbations in the condensate get exponentially amplified via parametric resonance, and then the perturbations re-scatter and the dynamics becomes highly nonlinearly. This is like a preheating scenario, and can be understood numerically via lattice simulations. By assuming spherical symmetry in Minkowski space, it was found that the Affleck-Dine condensate first decays into an excited Q-ball solution before settling to a stable Q-ball \cite{Enqvist:1999mv}. However, as we will see in this subsection, the Affleck-Dine fragmentation is actually more complex, and, in particular, CSQs are formed generically in this process.

To see this, we shall simulate the Affleck-Dine fragmentation in the early universe with an expanding lattice that follows an FRW evolution driven by the average energy density of the scalar field. (Fully numerical relativistic simulations are generally not needed, because scalar lumps are typically not sufficiently heavy to have strong self-gravity \cite{Kou:2019bbc, Kou:2021bij}.) There are currently quite a few mature codes to simulate preheating-like scenarios in a rigid FRW universe with scalar fields. In \cite{Zhou:2015yfa}, the Affleck-Dine fragmentation with a logarithmic potential has been simulated with the HLattice code \cite{Huang:2011gf} that uses a highly accurate sixth-order symplectic integrator. However, the focus in \cite{Zhou:2015yfa} was to study the production of the stochastic gravitational wave background in the model, which interestingly has a multiple peak structure that is linked to the sizes of the scalar lumps generated during the fragmentation. Since the gravitational wave production is only sensitive to the energy density changes in the fragmentation, the charge distribution was not investigated. Here we shall re-run the simulations in \cite{Zhou:2015yfa} and plot the charge density distributions in the fragmentation. 

\begin{figure}[tbp]
		\centering
		\includegraphics[width=0.3\textwidth]{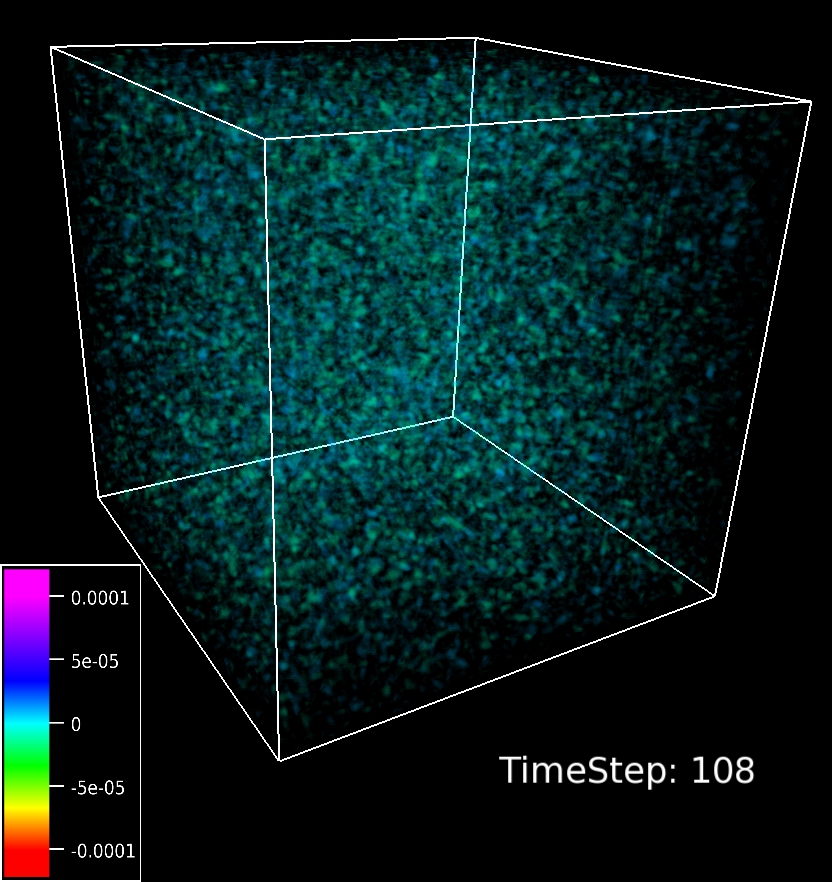}
		\hfill
		\includegraphics[width=0.3\textwidth]{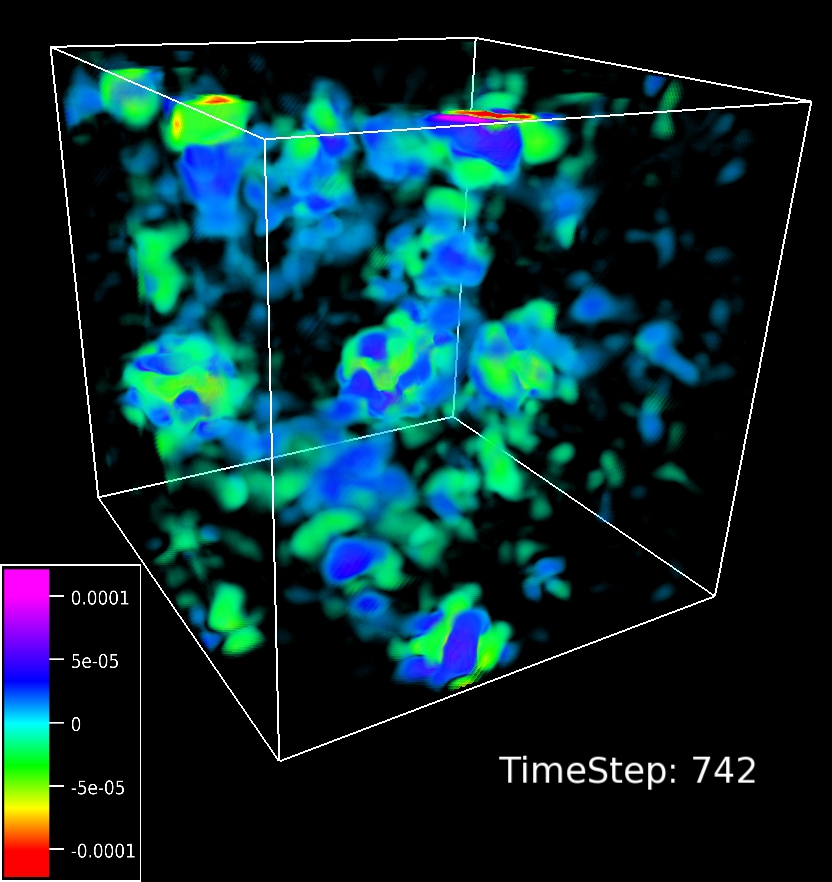}
		\hfill 
		\includegraphics[width=0.3\textwidth]{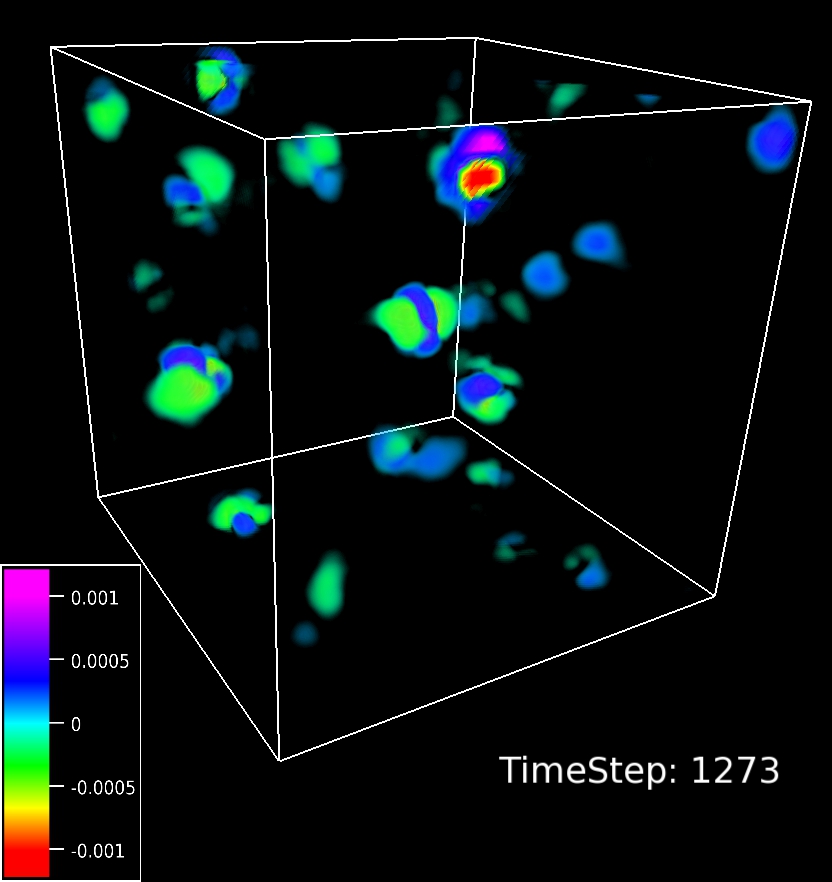}

        \includegraphics[width=0.3\textwidth]{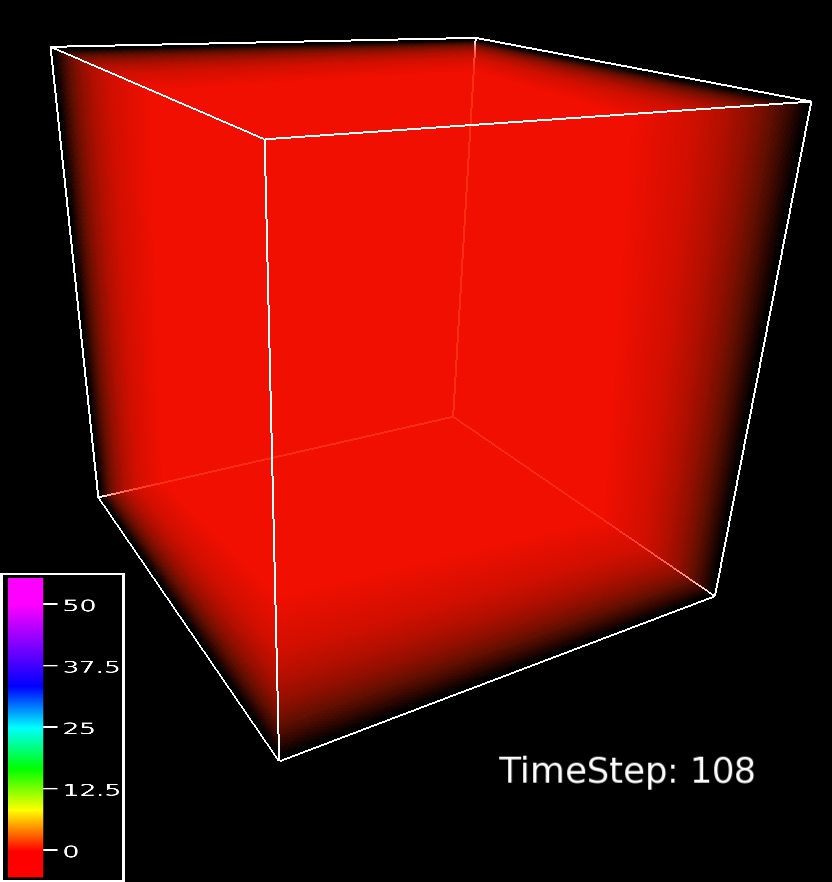}
		\hfill
        \includegraphics[width=0.3\textwidth]{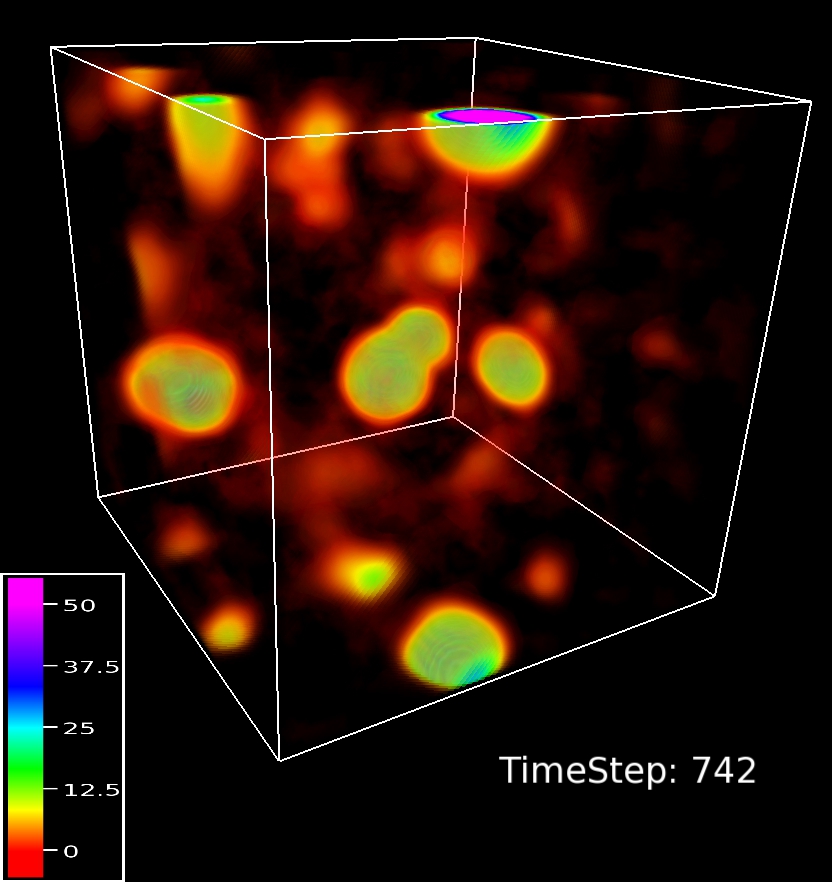}
		\hfill 
		\includegraphics[width=0.3\textwidth]{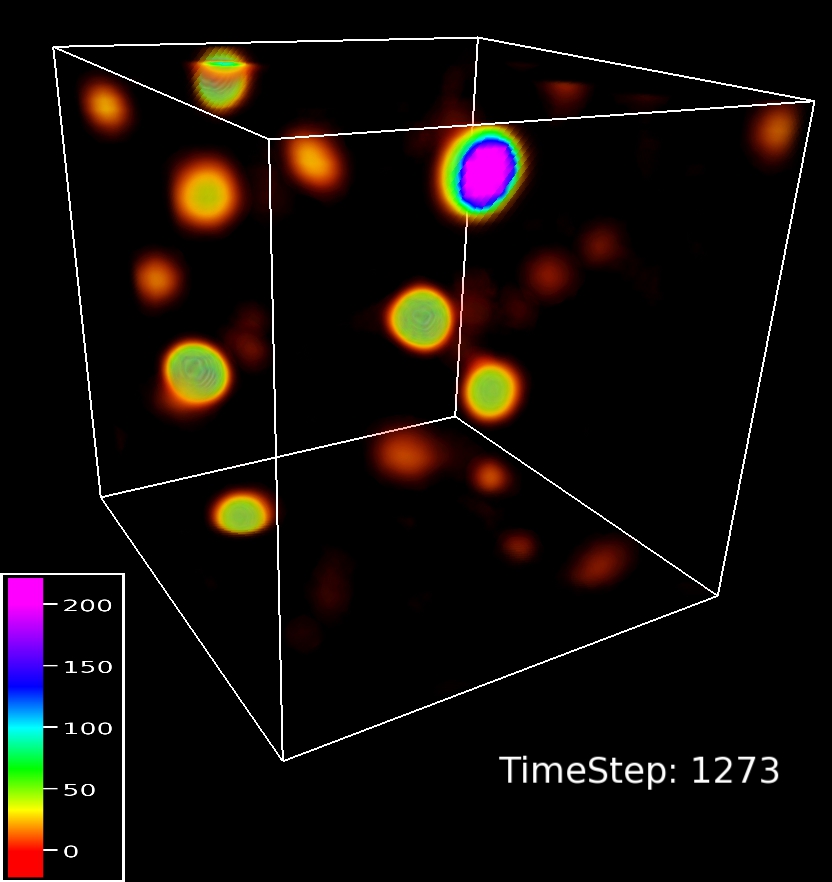}
		\caption{\label{fig:CSQsEU}
		Evolution of the charge density (top row) and energy density (bottom row) in an FRW universe with Gaussian random initial perturbations from inflation. We see that complex CSQs can naturally arise in the early universe with the logarithmic potential. Note that, for either the charge density or the energy density, the numerical scales for the color bars are different at TimeStep 1273: the numerical cutoff for the charge density is $10^{-4}$ at TimeStep 108 and 742 and is $10^{-3}$ at TimeStep 1273; the numerical cutoff for the energy density is 50 at TimeStep 108 and 742 and is 200 at TimeStep 1273. 1 TimeStep $\simeq 19.5/m$. The initial box size equals $0.05H^{-1}$, $H$ being the initial Hubble parameter.}
	\end{figure}

For concreteness, we shall consider the model where $K=-0.1$ and $m=1$TeV. The initial value for the static scalar condensate $\varphi$ is chosen to be $10^{16}$GeV, and the initial fluctuations of the condensate, which come from the classicalized quantum perturbations that exited the horizon during inflation and re-entered the horizon afterwards, are chosen to be Gaussian and of the size $|\delta \varphi / \varphi| = 10^{-5}$. We make use of a $256^3$ lattice with the built-in periodic boundary conditions, and the initial box size equals $0.05H^{-1}$, where $H$ is the initial Hubble parameter, and $dx = 10 dt \simeq 0.078/m$. In Figure \ref{fig:CSQsEU}, we have plotted the charge and energy density for three time instances. We find that a neutral Affleck-Dine condensate can fragment into positive and negative charges, which quickly form concentrated lumps, and the lumps are typically complex CSQs, rather than single charge (excited) Q-balls. Note that in the plots the numerical scales of the color bars vary at different time steps so as to visualize the perturbations at the time, and one TimeStep is equal to $250 dt \simeq 1.95/m$.

Motivated by the natural production of CSQs in the logarithmic potential from the Affleck-Dine fragmentation in the early universe, in the next section we will investigate the properties and long-term evolution of logarithmic CSQs. Our focus will be on the CSQs with relative low multipoles, as they are easier to construct and ideal to highlight the most salient properties of the logarithmic CSQs in simple ways. 	As we will see, these CSQs are extremely stable, so a periodic boundary condition, such as the one used in the HLattice code, would be unsuitable for long term evolutions of these CSQs, as perturbations emitted by the CSQ can bounce back and forth in the simulation box, polluting the determination of its lifetime. So in the next section, we shall re-write a new code with the LATfield2 framework \cite{Daverio:2015ryl} for these purposes, to adapt absorbing boundary conditions that are essential to evolve the system for an exceedingly long time.

	\section{Properties of logarithmic CSQs}
	\label{sec:stabiCSQs}
	
Having observed their potential role to play in the early universe, in this section we will take a more theoretical view to investigate the properties of CSQs in the logarithmic potential. Rather than generating them randomly in the early universe, we will now prepare them in a more controlled way, by simply superimposing opposite charge lumps, so as to expose their most salient features. We will see that the initial superimposed configuration will relax to the stationary CSQ phase without shedding away too much energy, unlike the $\varphi^6$ case \cite{Xie:2021glp}. To determine the long time stability of the logarithmic CSQs, we implement absorbing boundary conditions, which effectively prevent the radiation emitted from the CSQ from significantly perturbing and deforming the CSQ configuration. We find the logarithmic CSQs are extremely stable both in 3+1D and in 2+1D, and we have not observed their decay in our long term parallelized simulations. We also survey the parameter space of CSQ formation for different initial separations, velocities and oscillation frequencies of constituent lumps, and determine the ``attractor basin'' for these parameters.

		\begin{figure}[H]
		\centering
		\includegraphics[width=0.35\textwidth]{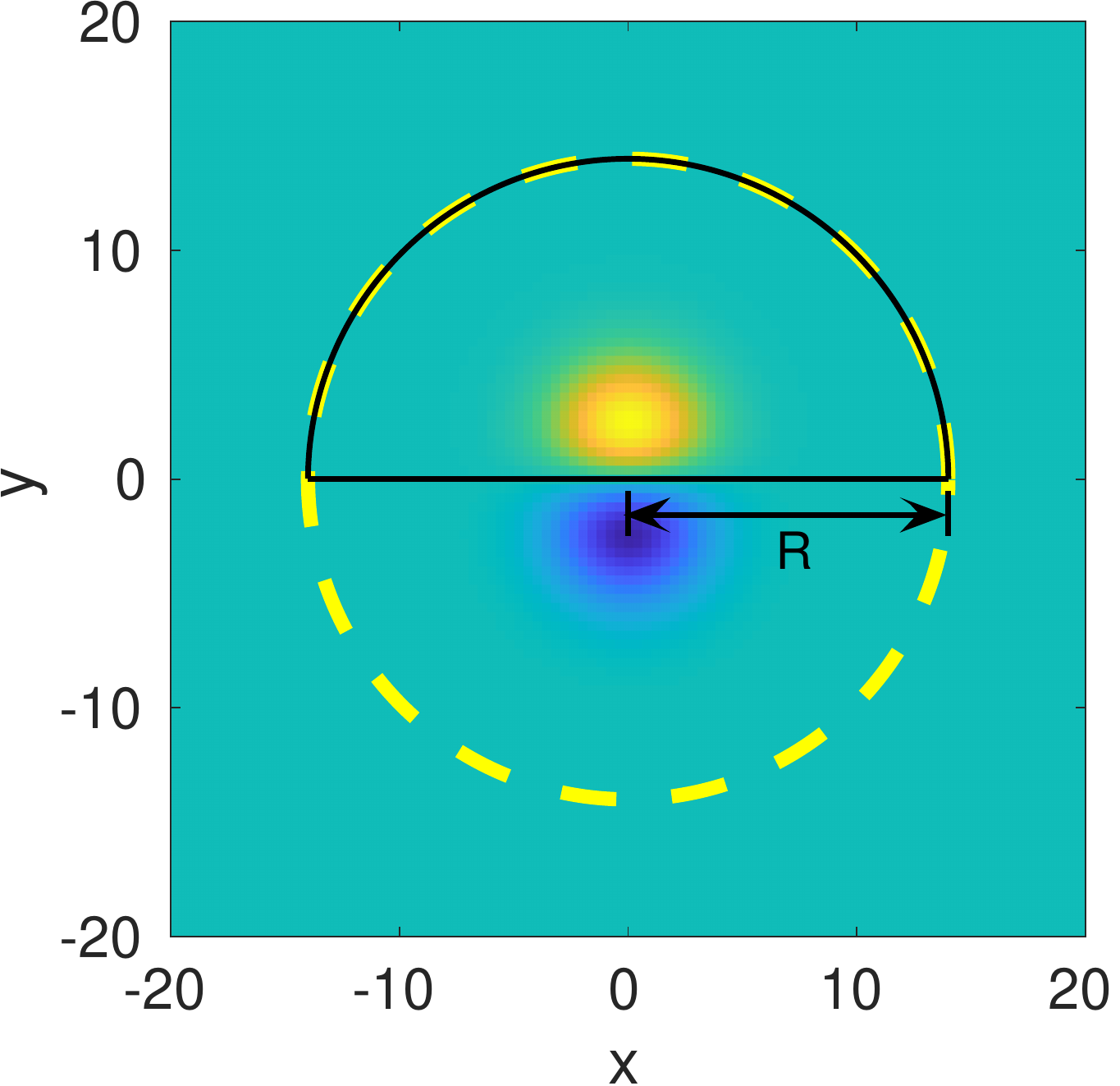}
		\caption{Regions (viewed on the $z=0$ slice) used to define $E$, $Q$, $E_c$, $Q_s$ $Q^{up}$ and $Q^+$.}
		\label{fig:regions}
          \end{figure}	

 To facilitate our later discussions about energy and charge integrations near the coordinate origin where the CSQ is placed, we define the following quantities:

	\begin{itemize}
		
		\item $E$ and $Q$: the total energy and total charge respectively in the simulation box 
		
		\item $Q^{up}$: the charge inside the upper half simulation box
		
                 \item $Q^+$: the charge obtained by integrating all positive charge densities over the simulation box 	
		\item $E_c$: the energy inside a sphere, which will be referred to as the ``central region'', with a radius of $14$ around the origin (yellow dashed circle in Figure \ref{fig:regions}); The value of $14$ is chosen so as to enclose almost all the energy and charge of the CSQ.
		
		\item $Q_s$: the charge inside the upper half of the sphere (solid black semi-circle in Figure \ref{fig:regions})
					
	\end{itemize}

	\subsection{Absorbing boundary condition}
	\label{sec:ABCs}

	The physical models concerned are in an infinite space. However, in our numerical simulations, we can only have a finite grid. Previously, the logarithmic model has been studied with a periodic boundary condition \cite{Copeland:2014qra}. While that approach is sufficient to establish the existence of CSQs as quasi-solitons, it would be insufficient to determine the long term stability of CSQs for two reasons. First, when we initially prepare the CSQs, we superimpose lumps of positive and negative charges (of which single Q-ball solutions would be one of the simplest choices) closely together and let the system relax to a CSQ. As CSQs are attractor solutions under the time evolution \cite{Copeland:2014qra, Xie:2021glp}, this is found to be an effective method to obtain CSQs. However, the relaxation process emits some amount of radiation. With a periodic boundary condition, the initial radiation would bounce back and forth in the box, which introduces destabilizing perturbations that are not in the physical system we are interested in. Second, as CSQs are quasi-stable, they emit further radiation as time goes by, which will also bounce within the box. Without reasonable methods to alleviate the unwanted radiation arising from the limitation of a lattice simulation, the long term stability of the CSQs can not be reliably established. This is especially true for the logarithmic potential, which, as we shall see, gives rise to extremely stable CSQs.

	\begin{figure}[tbp]
		\centering
		\includegraphics[width=0.45\textwidth]{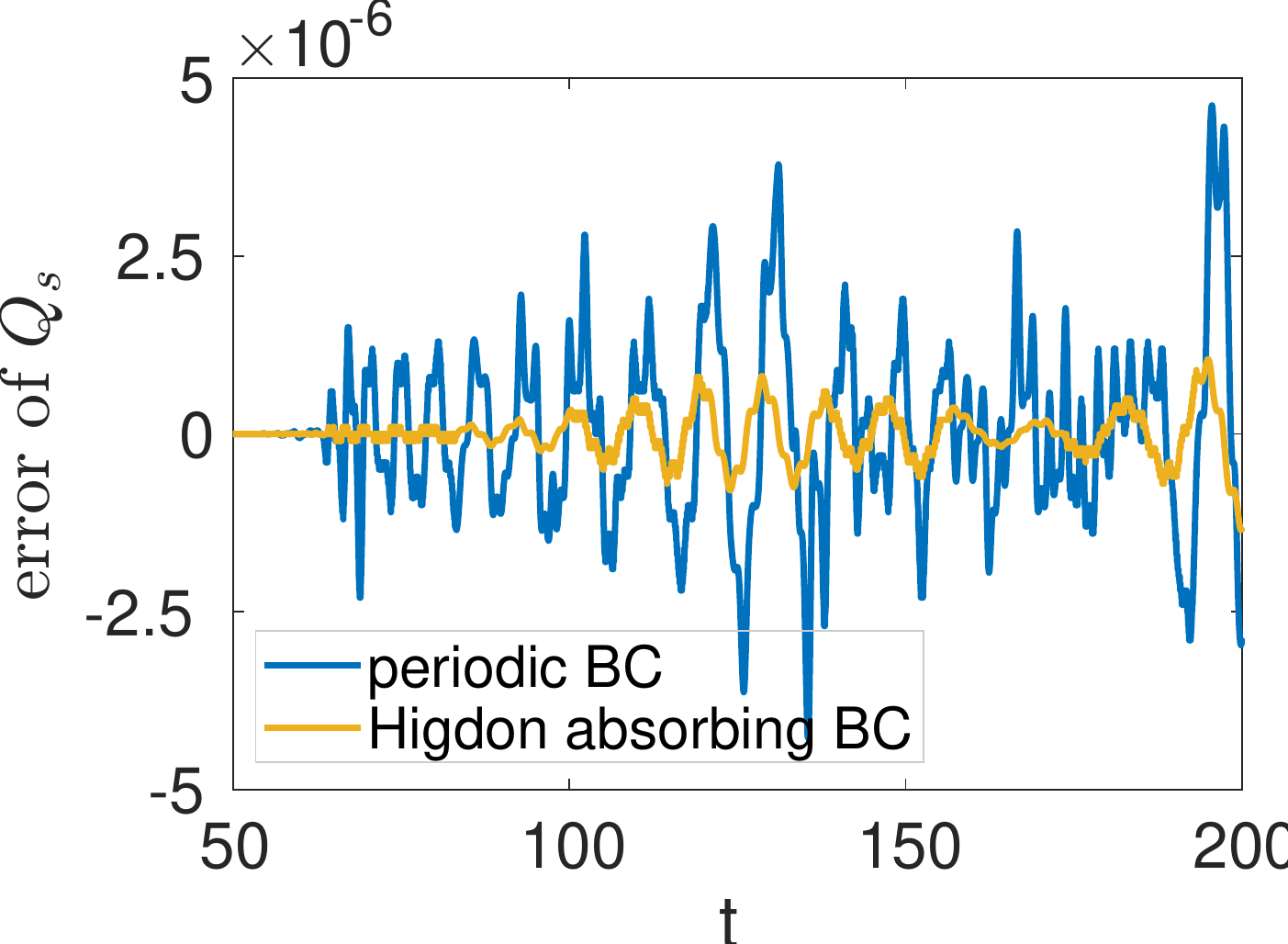}
		\caption{
		Evolution of the $Q_s$ charge (see text below Figure \ref{fig:regions} for its definition) for simulations with the periodic boundary condition and the 2nd order Higdon absorbing boundary condition. The ``error of $Q_s$'' is the error/deviation of $Q_s$ in the simulation with the periodic or absorbing boundary condition, as compared with a reference simulation where the waves have not propagated to the boundary of a larger simulation box at the end of the simulation.  
		}
	\label{fig:ABCs_effects}
	\end{figure}	
	
We shall make use of 2nd order Higdon's absorbing boundary conditions \cite{higdon1994radiation,higdon1986absorbing} for our simulation grid. Higdon's absorbing boundary conditions are designed based on the idea that a boundary condition that absorbs specific outgoing plane waves entirely will also absorb other plane waves partially and that we can use multiple of these conditions to improve the absorbing effect:
	\be 
	\prod_{j=1}^{M} \( \f{\pd}{\pd t}\pm c_j \f{\pd}{\pd x^i} \)\cphi|_{x^i=a}=0
	\ee 
where $a$ is the location of the boundary and the $c_j$'s are tunable constants that should be adjusted for particular problems. For example, the differential operator labelled by $j$ in the above formula can exactly absorb plane waves moving in the $x^i$ direction with a phase velocity equal to $c_j$, with the $+$ ($-$) sign for waves moving to larger (smaller) $x^i$ direction. Although higher order Higdon's methods have better absorbing effects, they are also more difficult to implement, due to the presence of higher order differential operators, and also computationally more expensive, and the second order Higdon's absorbing boundary condition is already rather effective for our purposes:  
	\bal
	\( \f{\pd^2}{\pd t^2}\pm (c_1+c_2)\f{\pd}{\pd x^i \pd t}+c_1 c_2\f{\pd^2}{\pd (x^i)^2} \)\cphi|_{x^i=a} &= 0 . \label{eq:Hig2}
	\eal
which will be the boundary condition we use later.

\begin{figure}[tbp]
		\centering
		\includegraphics[width=0.44\textwidth]{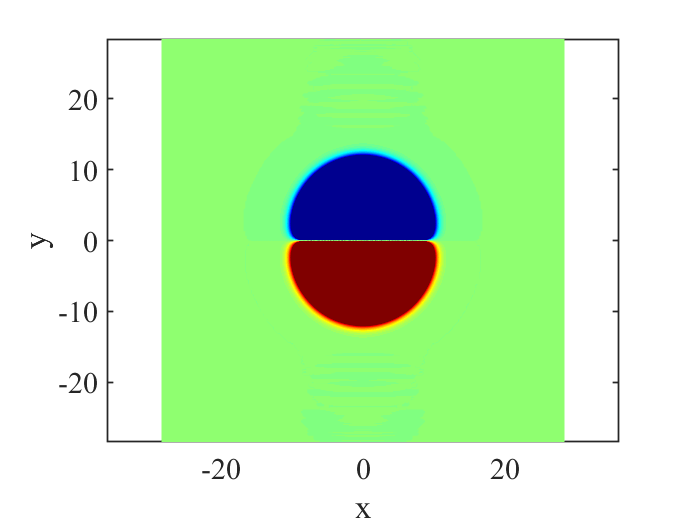}
		\includegraphics[width=0.44\textwidth]{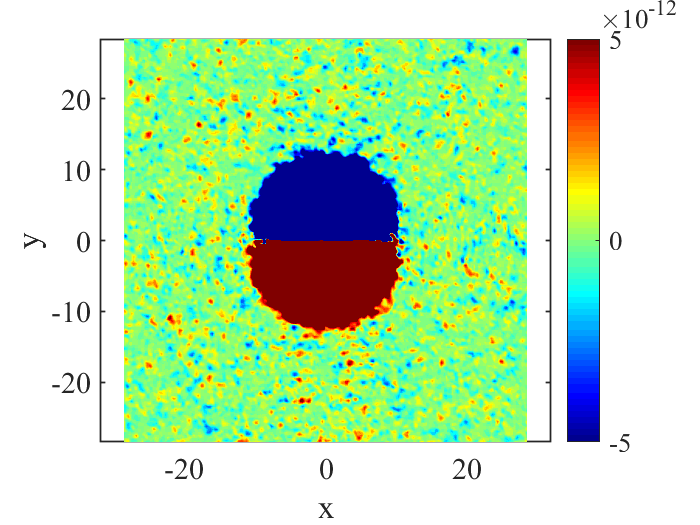}
		\hfill 
		\caption{\label{fig:perturbation}
Charge density at $t=2\times10^7$ with Higdon's 2nd order absorbing boundary condition (left plot) and at $t=1\times10^5$ with the periodic boundary condition (right plot) in a 2+1D simulation. We see that with a periodic boundary condition the CSQ configuration becomes quite noisy after a moderate amount of time. 
		}
	\end{figure}

To showcase the effectiveness of the 2nd order Higdon's absorbing boundary condition, in Figure \ref{fig:ABCs_effects}, we compare the initial evolution of $Q_s$, \ie the total charge in the $y\geq 0$ half central region (see the text below Figure \ref{fig:regions}), for three different boundary conditions: 1) periodic boundary condition; 2) 2nd order Higdon's absorbing boundary condition with $c_1=c_2=1$; 3) the reference simulation in which the lattice is chosen large enough such that the perturbations have not reached the boundary at the end of the test period, and which is taken to be the {\it bona fide} evolution. All the physical setup and other numeric parameters are chosen to be the same for the three evolutions. As we see in Figure \ref{fig:ABCs_effects}, in the beginning, the three simulations are the same as the perturbations have not propagated to the boundary; soon after, we can see that the periodic boundary condition and absorbing boundary condition deviate from the reference case, with the absorbing boundary much better than the periodic boundary. Additionally, we can see that the deviations from the reference case modulate with time, especially for the periodic boundary, which is due to the fact that the initial burst of radiation bounces back and forth in the simulation box and the CSQ is affected the most when the bounced radiation hits and passes through it. Note that, thanks to the ultra-soft interactions of the logarithmic potential, the deviations from the reference case is relatively small during the initial test period, even for the periodic boundary condition. However, in the long run, without proper absorption of radiation, the perturbations can accumulate and significantly affect the physical configurations. In Figure \ref{fig:perturbation}, we see that, after a time evolution of $2\times 10^7$, while the CSQ configuration still holds up very well for the Higdon's absorbing boundary condition, the periodic boundary condition introduces significant noise for the CSQ configuration just after a time evolution of $1\times 10^5$.
	
	\subsection{Numerical setup}

An interesting feature of the model with a logarithmic potential \eref{eq:Lagrangian} is that it has the Gaussian profile as its exact simple Q-ball solution \cite{bialynicki1975wave}: 
	\be
	\label{varphiini}
	\varphi(t,\bfx) = e^{\f{\oi^2 - 1}{2  K} + \f{D-1}{2}} e^{ K \f{|\bfx|^2}2} e^{i \oi t } ,
	\ee 
where $D$ is the number of the spatial dimensions and we have chosen the units according to \eref{eq:dimensionless quantities 1}. The $K$ parameter is chosen to be $-0.1$ through out this section.  We shall use this Gaussian profile as the constituent charge lumps, but it is worth emphasizing that this is not necessary, and other charged lumps can also be used in the construction. To get the simplest dipole CSQ configuration, we can superimpose two opposite charge lumps initially close to each other in symmetric locations along the $y$-axis in the center of the simulation box, and then let the configuration relax to the CSQ. The length of the simulation box is at least three times (often five times) that of the central CSQ region where most of the energy density is located within, depending on the parameters of the initial lumps. That is, we adjust the box size when the initial lumps become noticeably larger. If unstated, simulations in the following are performed in 3+1D. 	

\begin{figure}[tbp]
    \centering 
    \includegraphics[width=0.45\textwidth]{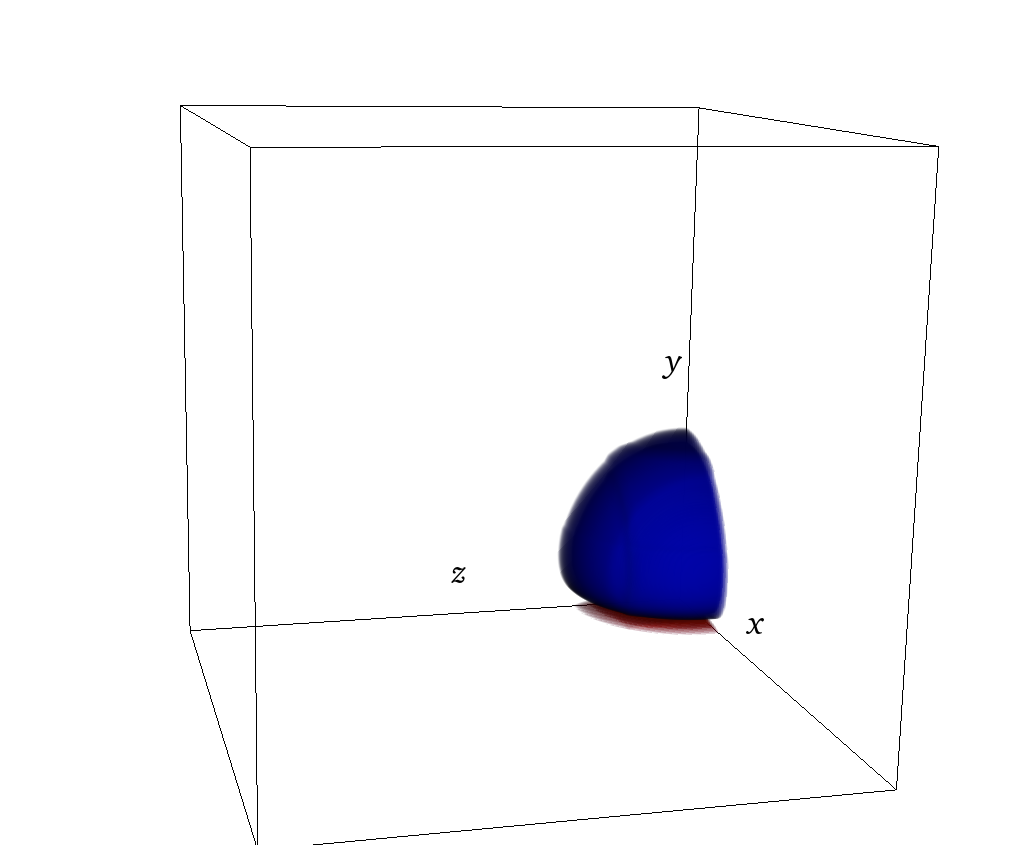}
    \caption{\label{fig:Schematic_diagram_of_symmetric_boundary_conditions}
    Thanks to the symmetry of a dipole (and as well as quadrupole) CSQ, we only simulate the first octant with $x,y,z>0$ and implement mirroring boundary conditions on the $x$-$y$, $y$-$z$ and $z$-$x$ planes, which provides a speed-up factor of 8.
    }
\end{figure}

Our code in this section makes use of the LATfield2 package \cite{Daverio:2015ryl} which provides a rudimentary framework for rapid development of parallel codes for classical lattice simulations, by defining objects such as lattice sites and fields on lattice. The Runge-Kutta 4th order method is used to evolve the system in time and spatial derivatives are approximated by 4th order finite differences. As mentioned, we use a 2nd order Higdon's absorbing boundary condition to absorb out-going waves to reduce unphysical in-going reflection of waves, with the absorbing parameters chosen to be $c_1 = c_2 =1$. As the CSQ is prepared in the origin and along the $y$ axis, for a dipole CSQ, the field is symmetric with respect to the $x$-$y$ and $y$-$z$ plane, and its real (imaginary) component is symmetric (anti-symmetric) with respect to the $z$-$x$ plane. To speed up the simulations on top of the parallelization, we only simulate the first octant with positive $x,y,z$ and implement mirroring boundary conditions on the $x$-$y$, $y$-$z$ and $z$-$x$ planes; see Figure \ref{fig:Schematic_diagram_of_symmetric_boundary_conditions}. This provides a speed-up of a factor of 8 for 3+1D simulations and a factor 4 for 2+1D simulations, where we simulate the first quadrant. Similar arrangements can also be made for higher multipole CSQs such as the quadrupole one. 

A numerical subtlety when preparing the initial lumps is that the potential \eref{eq:log potential} contains a logarithm ${\rm ln}|\varphi|^2$. While the potential itself is not divergent at $|\varphi|=0$, thanks to the $|\cphi|^2$ factor in front, numerically, we may encounter problems, especially with the exponential profile \eref{varphiini}, which gives rise to small numbers due to the $e^{K {|\bfx|^2}/2}$ factor with negative $K$ and large $|\bfx|$, exceeding the double precision. To overcome this, a small regulator $\cepi$ is added in, ${\rm ln}(|\varphi|^2+\cepi)$, when numerically evolving the equations of motion, with $\cepi$ chosen to be $1 \times 10^{-60}$ times smaller than the maximum value of the $|\cphi|$ field. We have varied this ratio from $1 \times 10^{-70}$ to $1\times 10^{-50}$ and found no noticeable differences in the simulation results.

	\subsection{Stability of CSQs}
	\label{sec:evolution_of_CSQs}

	\begin{figure}[tbp]
		\begin{subfigure}{0.22\textwidth}
			\caption{$t=2$}
			\includegraphics[width=\textwidth,trim=160 100 100 150,clip]{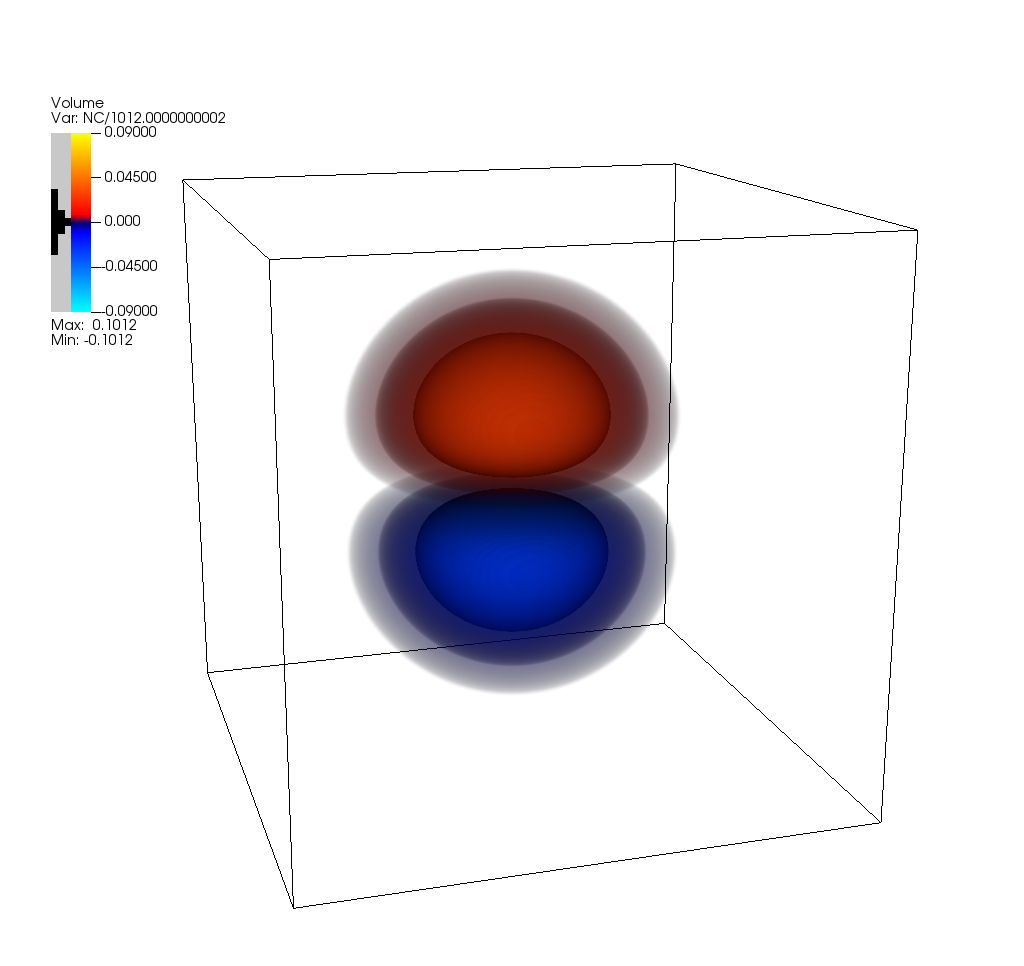}
		\end{subfigure}
		\hfill 
		\begin{subfigure}{0.22\textwidth}
			\caption{$t=12$}
			\includegraphics[width=\textwidth,trim=160 100 100 150,clip]{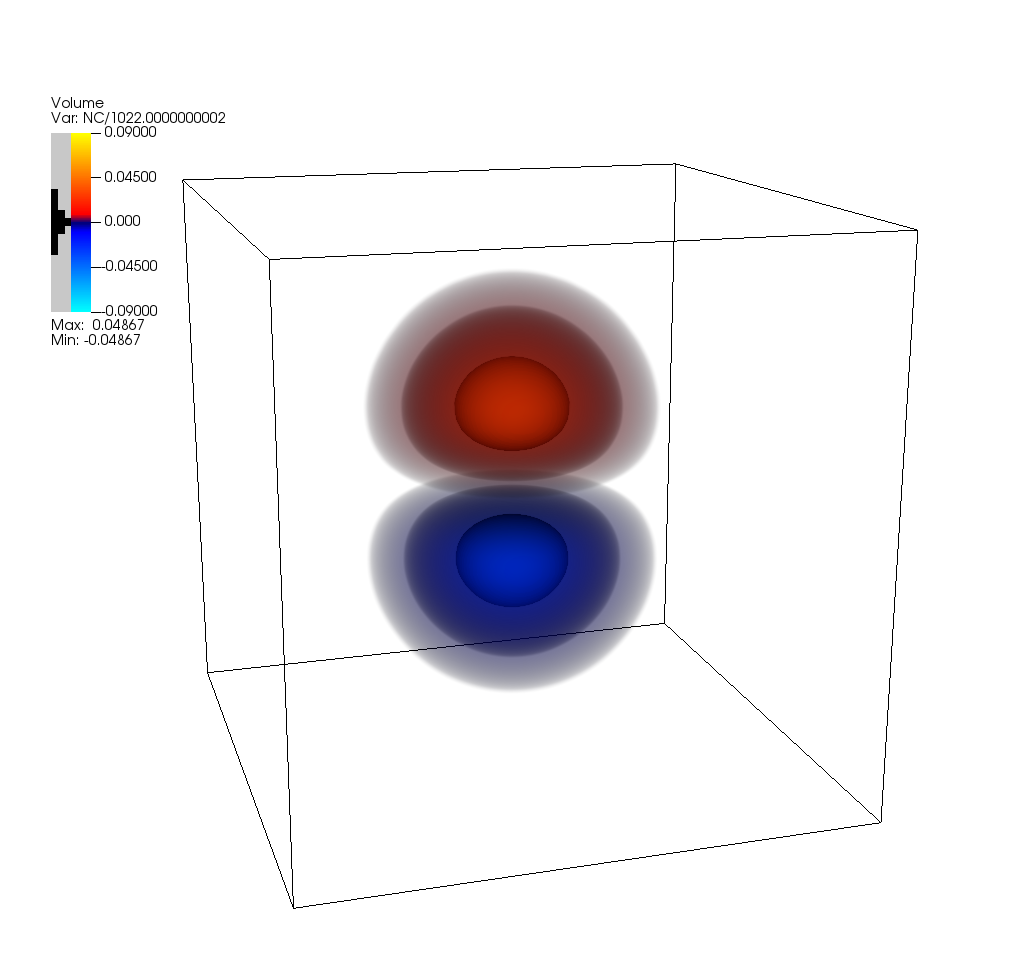}
		\end{subfigure}
		\hfill 
		\begin{subfigure}{0.22\textwidth}
			\caption{$t=22$}
			\includegraphics[width=\textwidth,trim=160 100 100 150,clip]{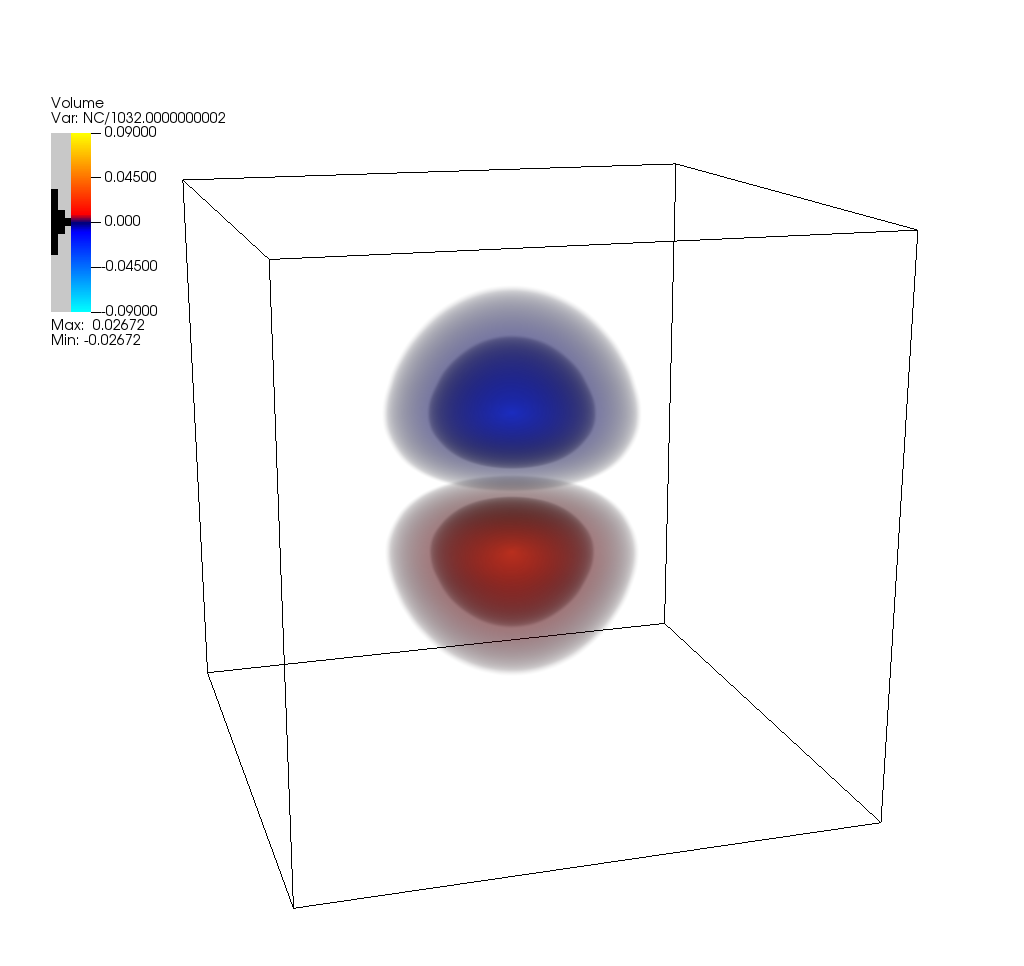}
		\end{subfigure}
		\hfill 
		\begin{subfigure}{0.22\textwidth}
			\caption{$t=32$}
			\includegraphics[width=\textwidth,trim=160 100 100 150,clip]{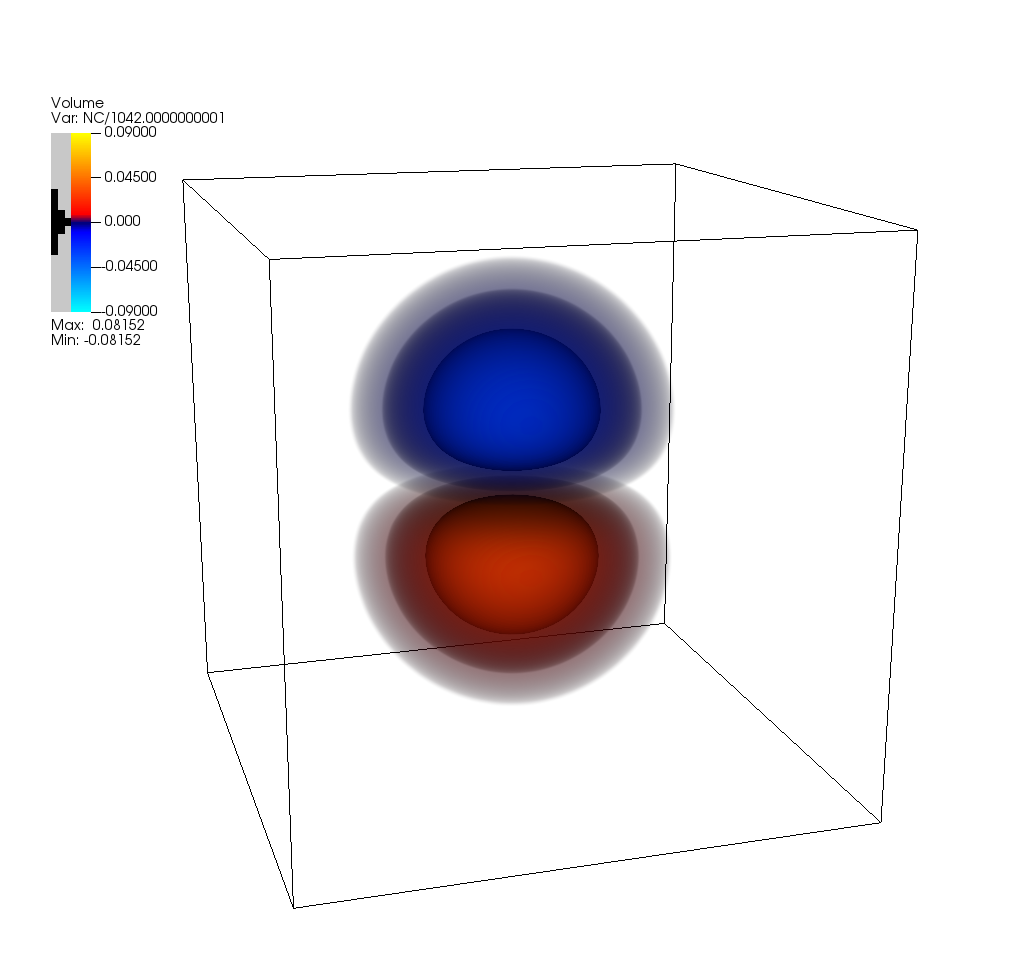}
		\end{subfigure}
		\begin{subfigure}{0.22\textwidth}
			\caption{$t=42$}
			\includegraphics[width=\textwidth,trim=160 100 100 150,clip]{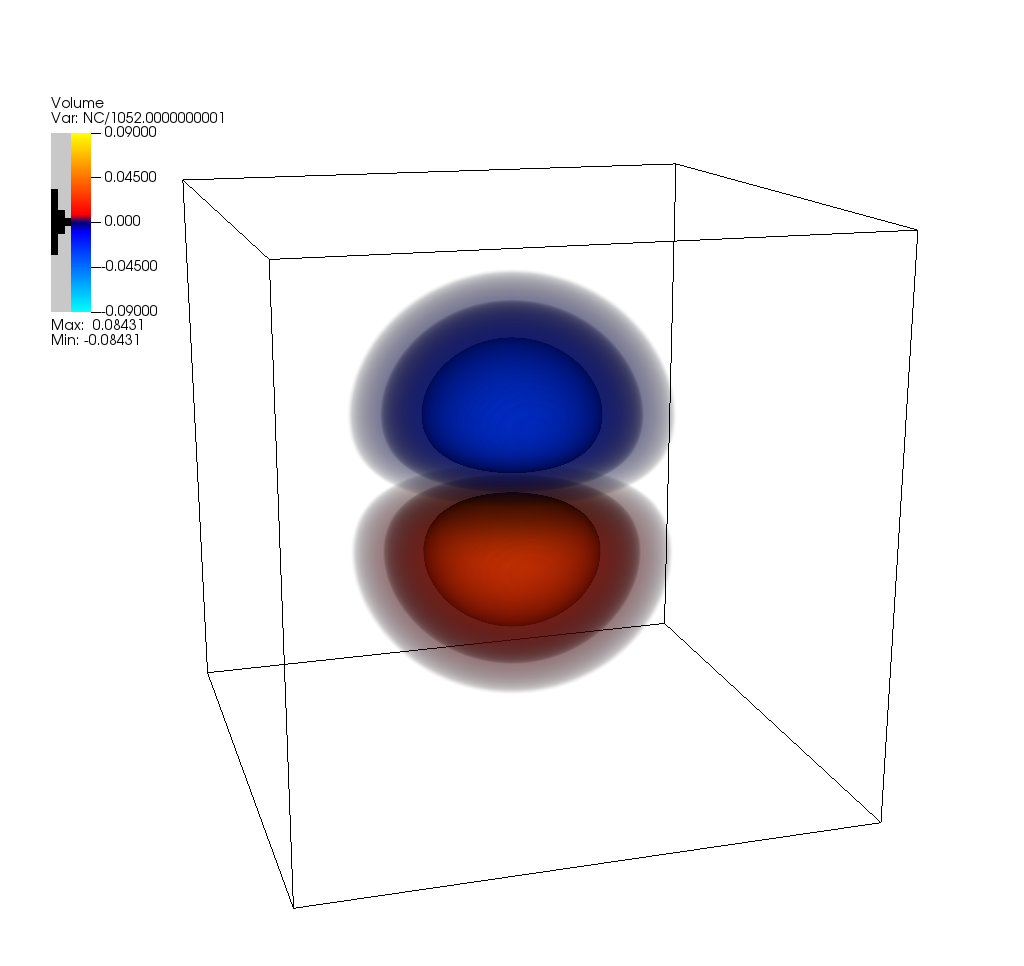}
		\end{subfigure}
		\hfill 
		\begin{subfigure}{0.22\textwidth}
			\caption{$t=52$}
			\includegraphics[width=\textwidth,trim=160 100 100 150,clip]{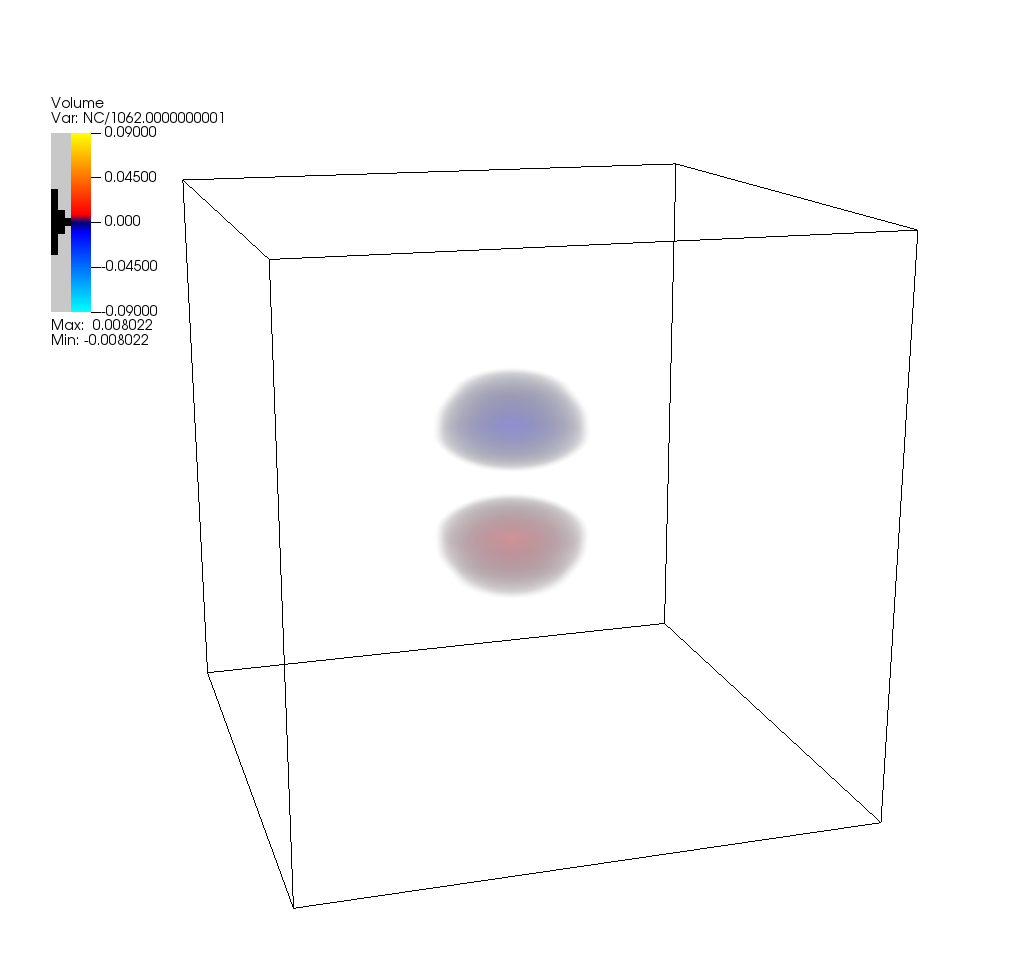}
		\end{subfigure}
		\hfill 
		\begin{subfigure}{0.22\textwidth}
			\caption{$t=62$}
			\includegraphics[width=\textwidth,trim=160 100 100 150,clip]{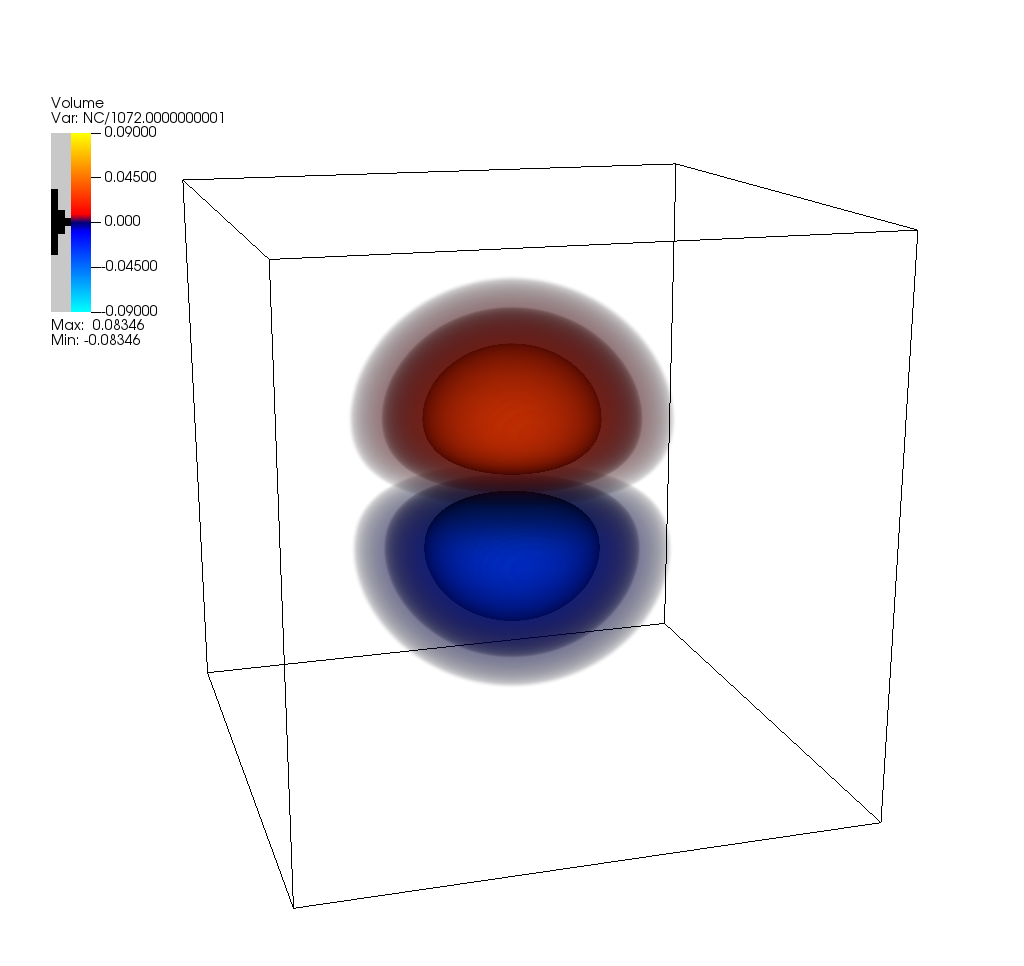}
		\end{subfigure}
		\hfill 
		\begin{subfigure}{0.22\textwidth}
			\caption{$t=72$}
			\includegraphics[width=\textwidth,trim=160 100 100 150,clip]{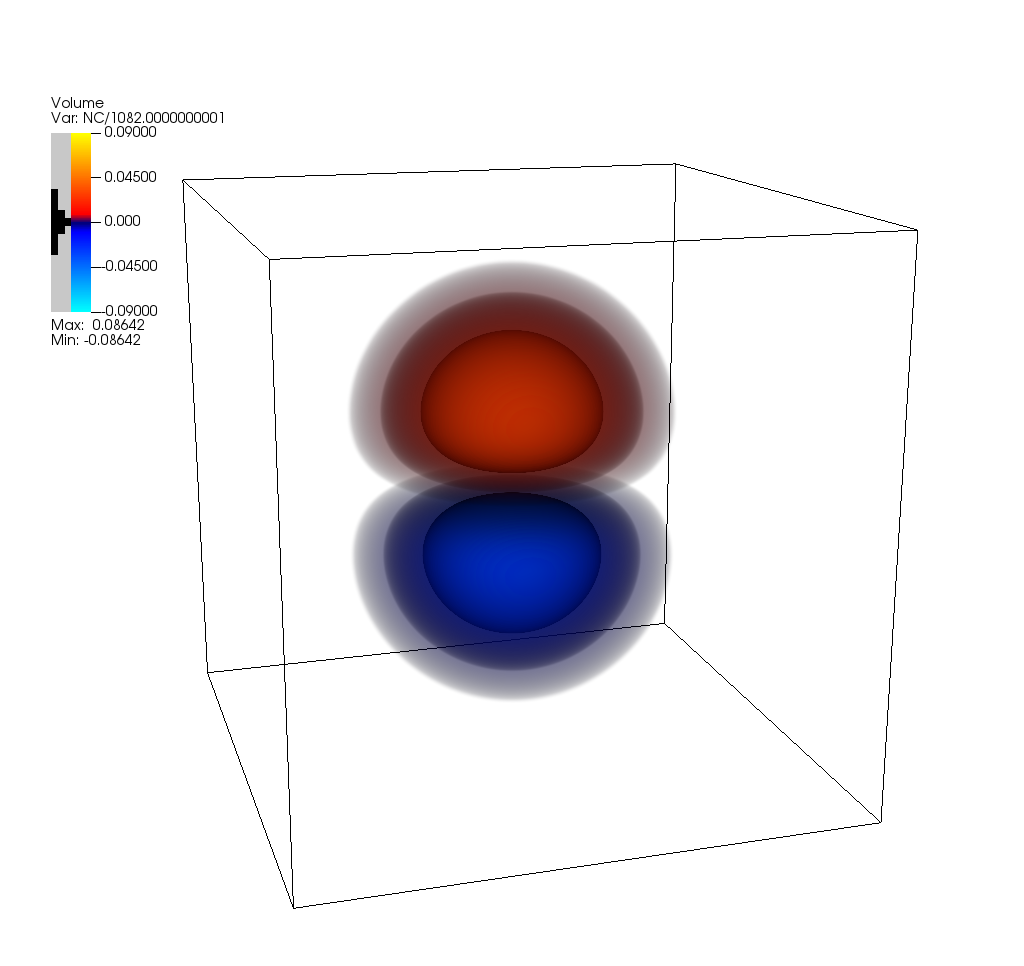}
		\end{subfigure}
		\caption{\label{fig:evolution_sequence_of_ordinary_CSQs}
			Evolution sequence of the charge density of a dipole CSQ. The energy density, which is not plotted, remains mostly spherically symmetric. The red and blue color denote positive and negative charges respectively. This sequence is extracted after the CSQ has evolved for a period of $1000$. The plots cover about $1/8$ of the whole 3D lattice. 
		}
	\end{figure}

As we have shown, CSQs may be formed in the early universe from fragmentation of some VEV in a preheating-like process. For this reason, we will focus on the 3+1D simulations in this subsection. As a fiducial example, we will investigate the properties of a dipole CSQ evolved from superimposing a Q-ball with $\oi = 1.2$ and an anti-Q-ball with $\oi = -1.2$ (cf.~\eref{varphiini}), with an initial separation between their centers equal to $2$. We will see that the properties of CSQs in the logarithmic potential are qualitatively similar to that in the $\varphi^6$ potential  \cite{Xie:2021glp}, but they are much more stable.
	
\begin{figure}[tbp]
		\centering 
		\includegraphics[width=0.45\textwidth]{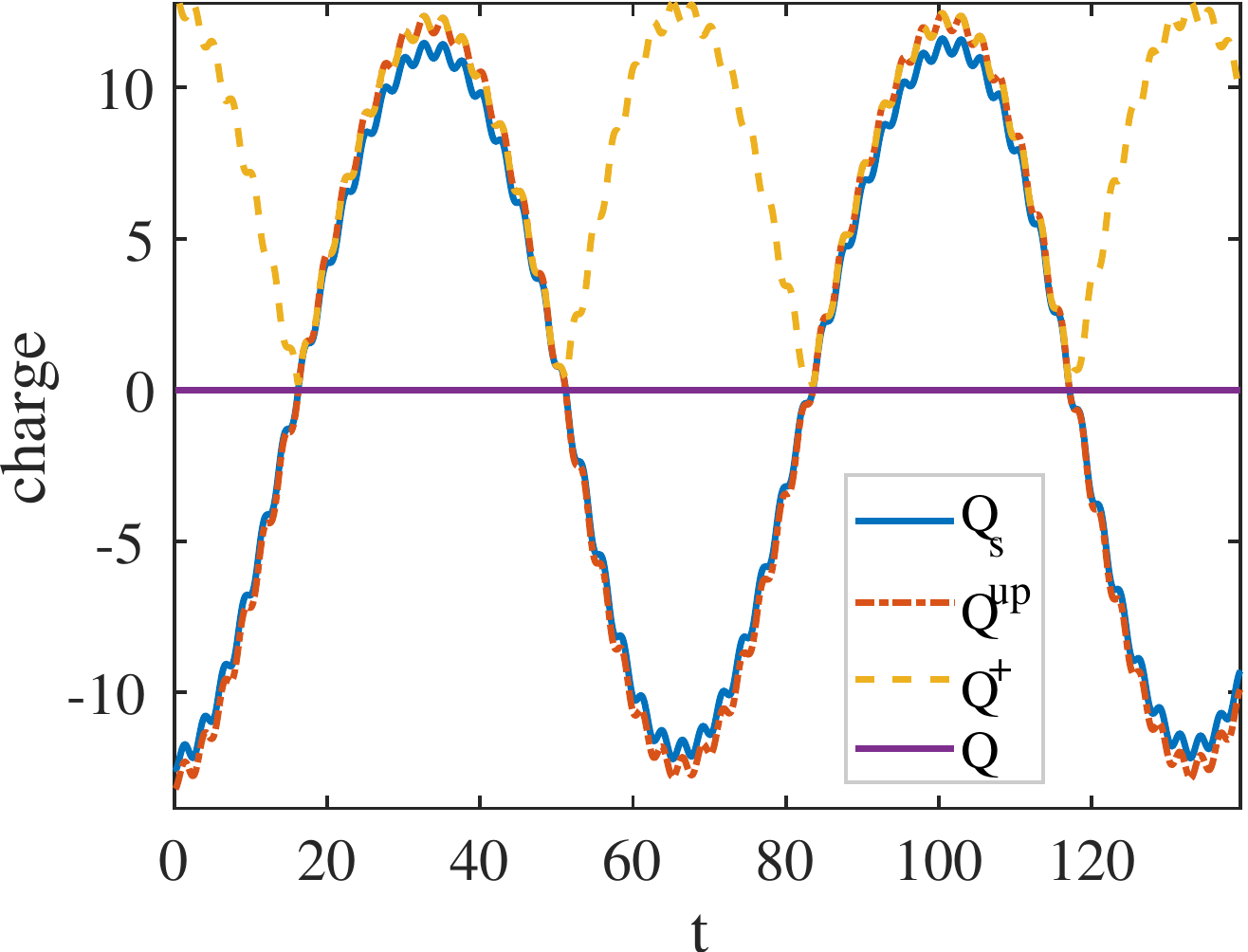}
		\caption{\label{fig:short_time_charge_evolution_of_ordinary_CSQs}
			Evolution of various charges. See Figure \ref{fig:regions} and the text below for the definitions of the quantities $Q_s$, $Q^{up}$, $Q^+$ and $Q$. 
		}
	\end{figure}

First, let us visualize the charge-swapping behavior of the CSQ in more details. In Figure \ref{fig:evolution_sequence_of_ordinary_CSQs}, we plot the evolution sequence of the charge density of the dipole CSQ in about one charge-swapping period, that is, the process of the positive charge lump becoming negative and then turning back to being positive. In Figure \ref{fig:short_time_charge_evolution_of_ordinary_CSQs}, we also plot the evolution of various charges defined in the text below Figure \ref{fig:regions}. The purple line is the total charge of the whole region $Q$ and is always zero as expected, since the two initial constituent Q-balls have opposite charges. The red dot-dashed line, denoted by $Q^{up}$, is the total charge obtained by integrating over the up half simulation box, while the blue solid line, denoted by $Q_s$, which closely trails the red dot-dashed line, is the total charge obtained by integrating over the half spherical region defined in Figure \ref{fig:regions}. The dashed yellow line, denoted by $Q^+$, is the total charge obtained by summing all the positive charges in the simulation box. We see that most of the charges are within the CSQ, and the charge swapping period of the CSQ is much longer than the oscillation period of the field, the latter of which gives rise to the small wiggles on the large modulation that is mostly sinusoidal.  We emphasize that while the charge density of the CSQ is dipolar and swapping, its energy density is approximately spherically symmetric.

\begin{figure}[tbp]
		\centering
		\includegraphics[width=0.48\textwidth]{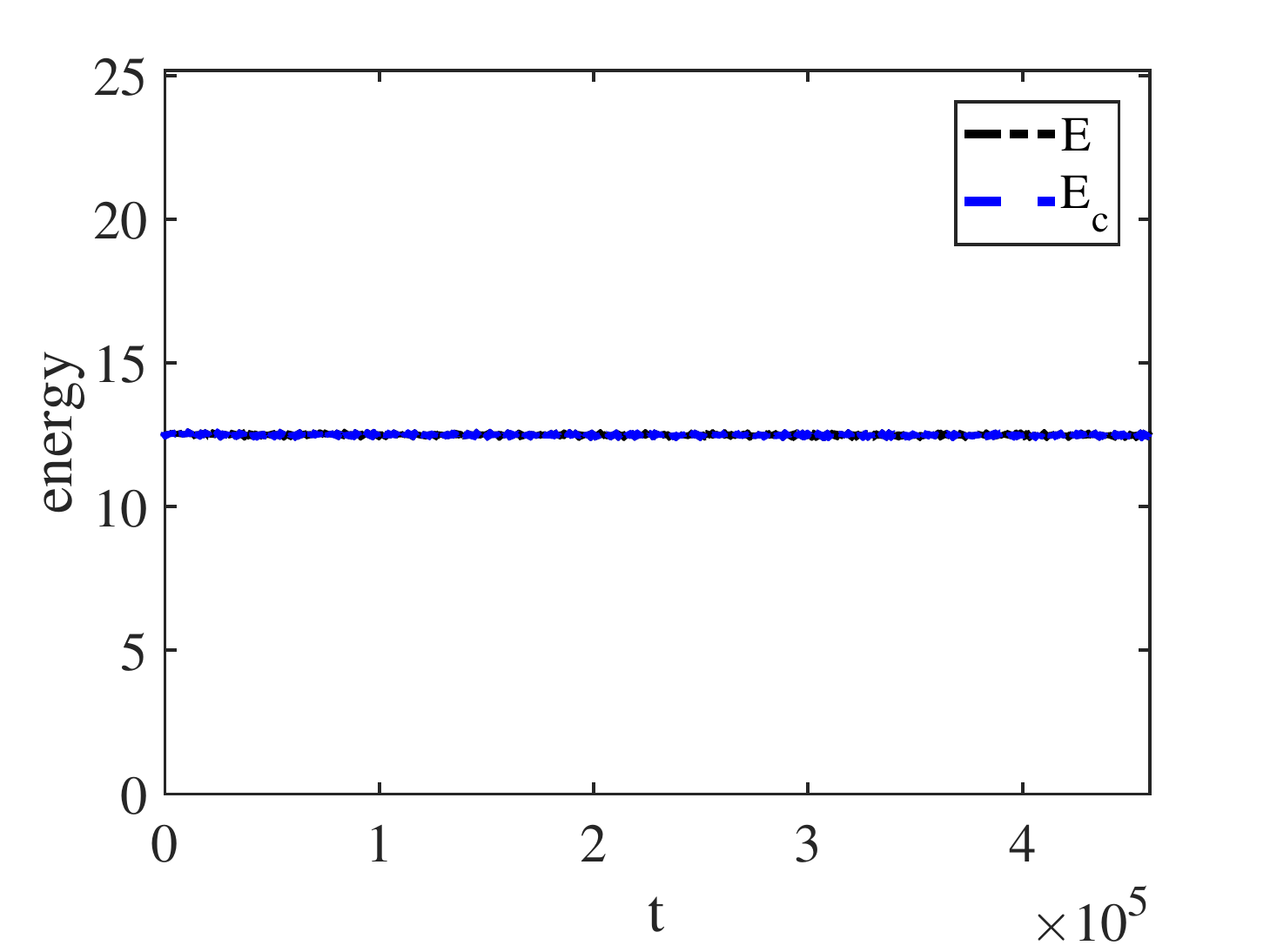}
		\hfill 
		\includegraphics[width=0.45\textwidth]{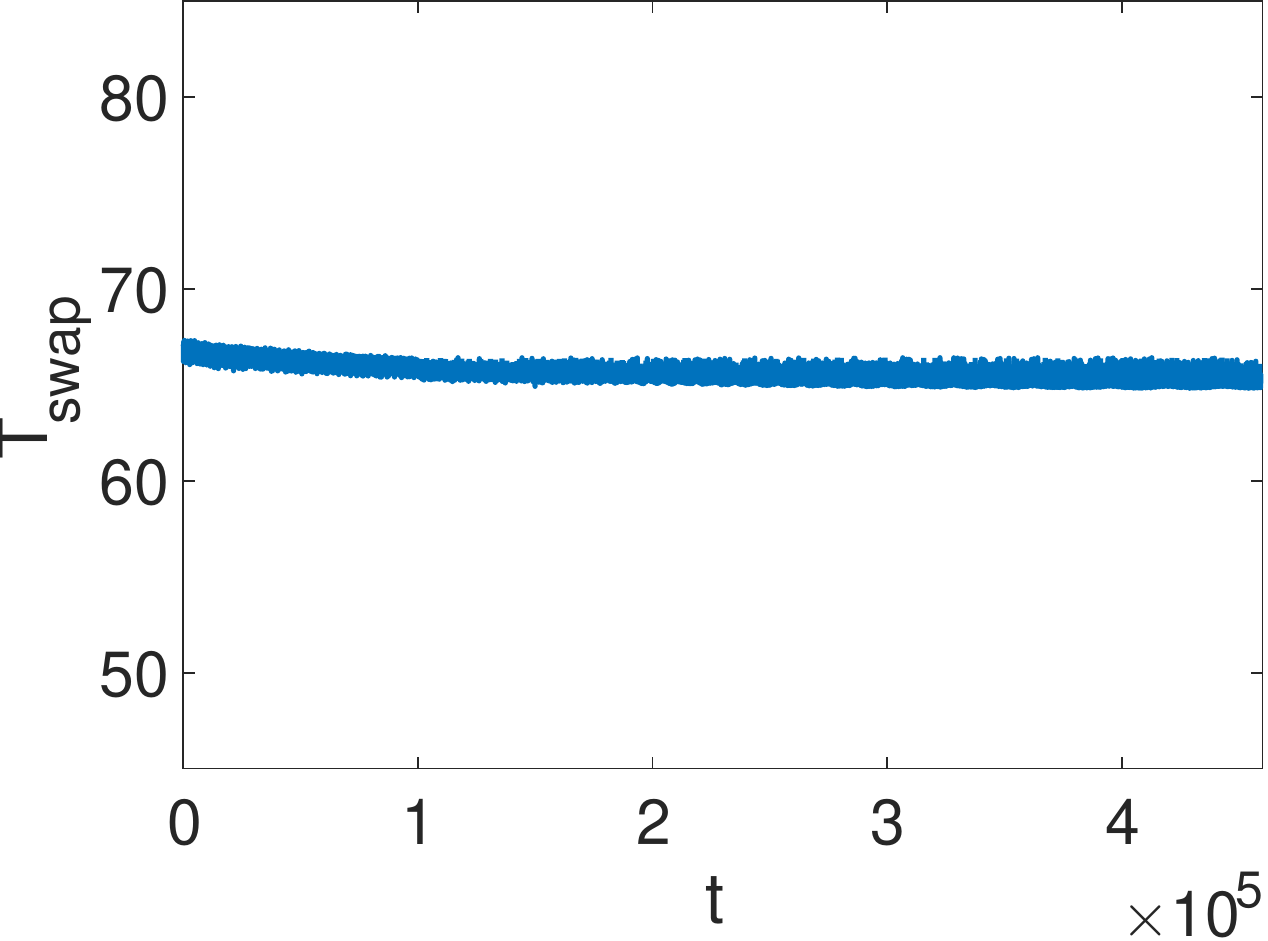}
		\caption{\label{fig:long_time_evolution_energy_and_period}
			Long term evolution of the energy and charge-swapping period of the logarithmic CSQ in 3+1D. The black line in the left plot, which is essentially invisible, is the total energy obtained by integrating the whole simulation box, while the blue line, denoted by $E_c$, is the energy of the central region (see Figure \ref{fig:regions}). Note that since we are only simulating the first octant, the energy in the left plot in only $1/8$ of the CSQ's total energy. This is achieved by about $1.1 \times 10^{5}$ CPU hours.
		}
	\end{figure}

The ultra-solfness of the logarithmic potential means that its relaxation process is quite different from that in the $\varphi^6$ potential \cite{Xie:2021glp}. For the $\varphi^6$ case, after superimposing two charge lumps, the initial configuration quickly sheds away a significant amount of energy and settles down to the CSQ plateau. For the logarithmic case, however, we can barely see a relaxation process, and only a very small amount of energy ebbs away in the beginning. Also, during the relaxation, the lump in the $\varphi^6$ case does not have a well-defined swapping frequency. However, the lump in the logarithmic case starts to swap charges with a well-defined changing swapping frequency (within a relatively broad band) from the beginning, and the charge swapping frequency changes slowly for a relatively long time before settling down to a constant band. More interestingly, due to the ultra-solfness of the potential, the logarithmic CSQ is extremely stable. (This is of course when a CSQ can stably form from certain initial conditions, such as the ones we use in this subsection; outside the attractor basin of a CSQ, a swapping lump disintegrates very quickly; see Section \ref{sec:attractorBasin}.)  Despite the acceleration of the simulation with parallelization and only evolving the octant, we have not captured the decay of the logarithmic dipole CSQs in 3+1D or in 2+1D. In Figure \ref{fig:long_time_evolution_energy_and_period}, we plot the evolution of energies and the charge-swapping period of the dipole CSQ. We see that in 3+1D the logarithmic CSQ's lifetime is longer than
\be
{(\rm lifetime)}^{\rm 3+1D}_{\rm logCSQ} \gtrsim 4.6\times 10^{5}/m .
\ee
To achieve such a long simulation with a $256^3$ lattice for the first octant, we have used about $1.1 \times 10^{5}$ CPU hours. In 2+1D, the logarithmic dipole CSQs have similar properties, and their lifetimes are found to be longer than 
\be
{(\rm lifetime)}^{\rm 2+1D}_{\rm logCSQ} \gtrsim 2.5\times 10^{7}/m ,
\ee
which is achieved by about $2 \times 10^4$ CPU hours. Furthermore, we have also checked the stability of the quadrupole ones, and found that they are also stable to a similar time scales; Specifically, 
in 2+1D, the quadrupole logarithmic CSQ's lifetime is longer than $2.3\times 10^{7}/m$, which takes $1.8\times 10^{4} $ CPU hours to simulate. We emphasize that for such long simulations to be reliable, it is essential to utilize absorbing boundary conditions to absorbing outgoing radiation at the boundary of the simulation box.

\begin{figure}[tbp]
		\centering
		\includegraphics[width=0.45\textwidth]{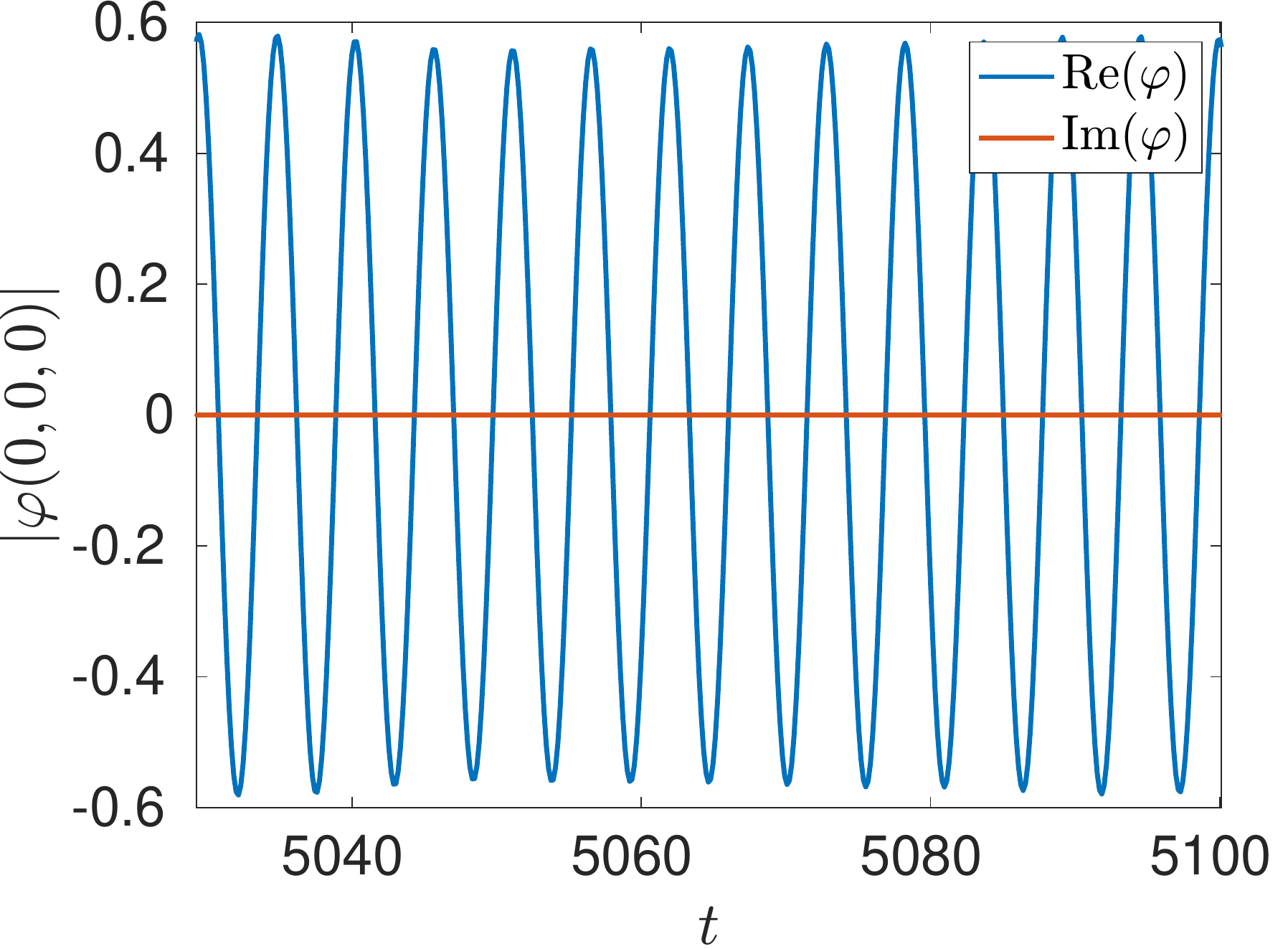}
		\hfill 
		\includegraphics[width=0.45\textwidth]{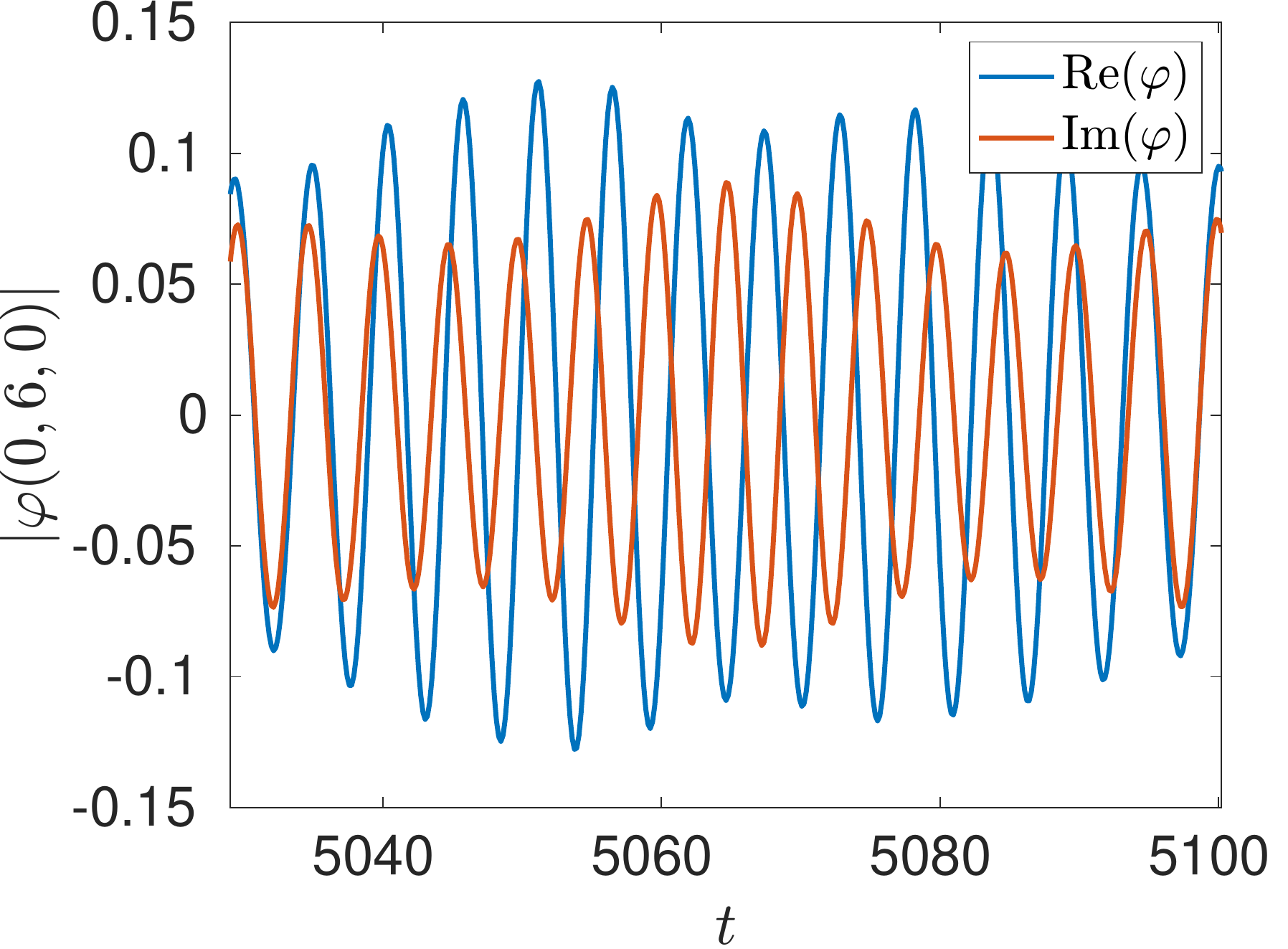}
		\includegraphics[width=0.45\textwidth]{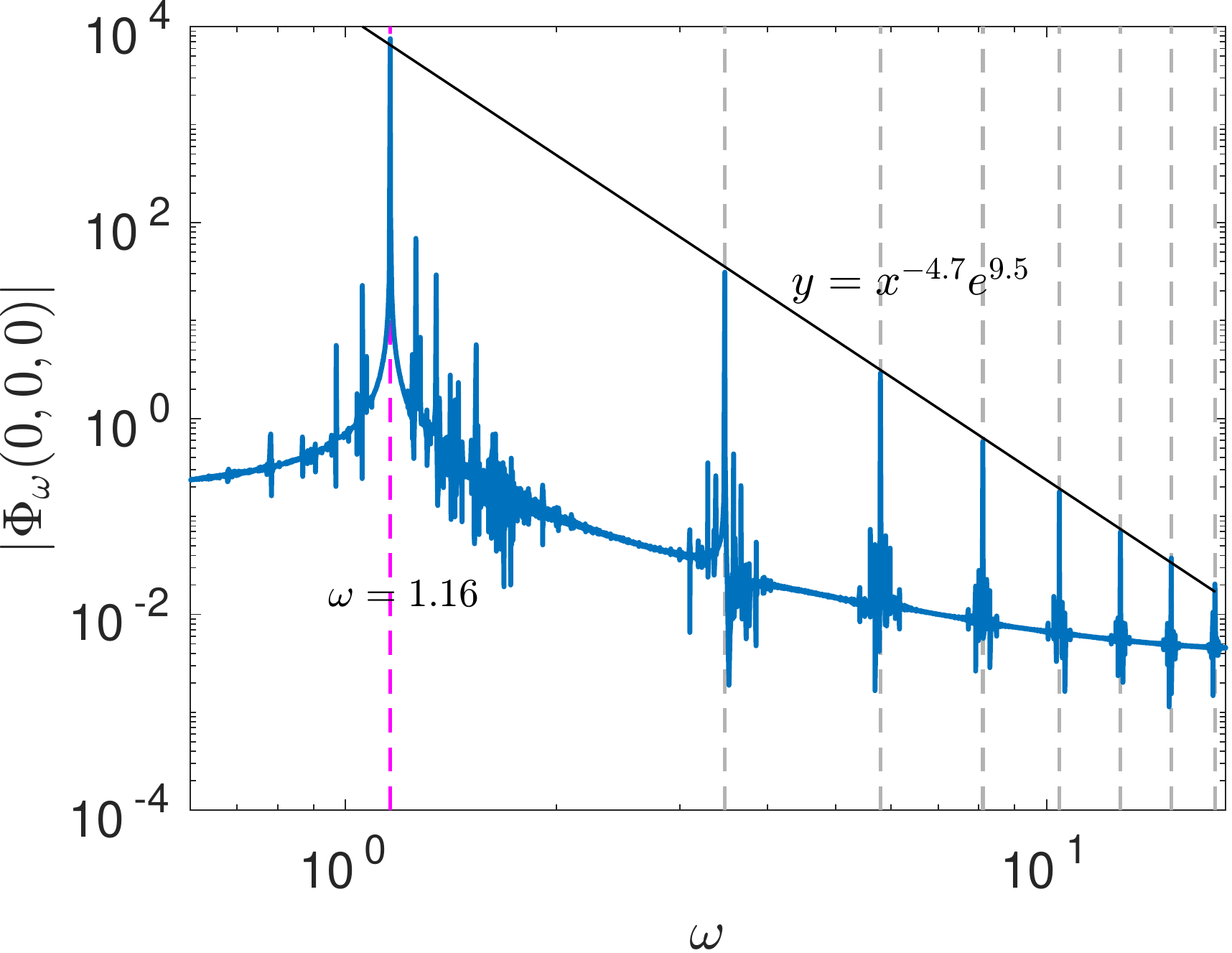}
		\hfill 
		\includegraphics[width=0.45\textwidth]{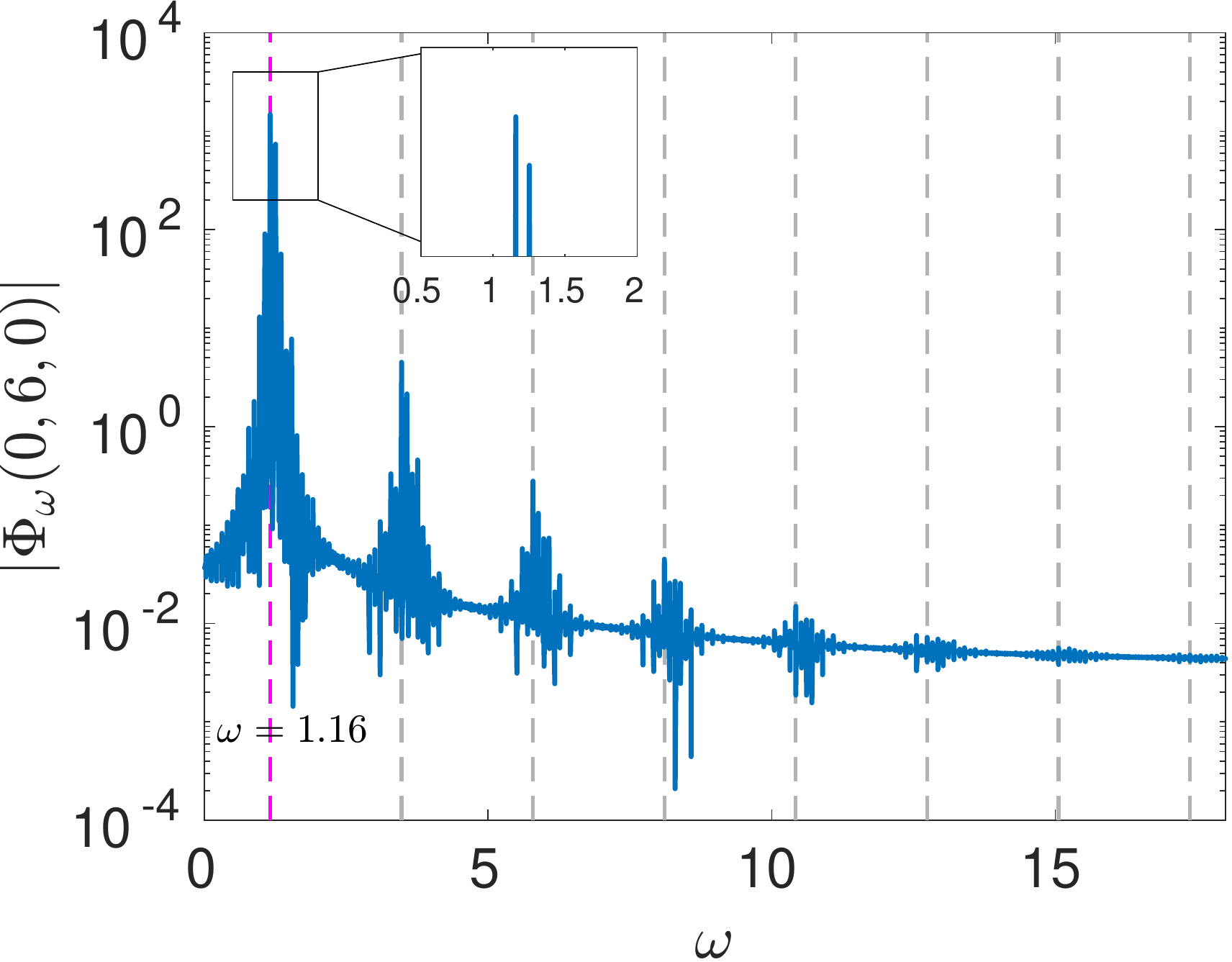}
		\caption{\label{fig:time_fourier_of_CSQs}
			Evolution of the field and its Fourier spectrum at two spatial points. The left column is at the center of the CSQ and the right column is at a point in the outer region of the CSQ. In the Fourier spectrum plots, the vertical dashed magenta lines show the base frequency and the vertical dashed gray lines show the odd multiples of the base frequency. In the inset of the bottom right plot, we can see two sub-peaks for the dominant peak, the difference between the two sub-peaks being the charge-swapping frequency of the CSQ. 
		}
	\end{figure}

It is instructive to see the spectra of the oscillations of the CSQ. To this end, we pick two representative points, and plot the evolution of the scalar field at these points and their corresponding Fourier spectra in Figure \ref{fig:time_fourier_of_CSQs}. In the top left plot, the field evolution at the origin, \ie  the center of the CSQ, has been plotted, which has an oscillating real part with a frequency larger than the charge-swapping frequency and a vanishing imaginary part due to the dipole symmetry of the CSQ. The bottom left plot shows the Fourier spectrum of this point's time evolution for the time from $t=0$ to $t \sim 2.7 \times 10^4$. We see that  the peaks of the spectrum are odd multiples of a base frequency $\oi \approx 1.16$, and decrease according to  the power law of $e^{9.5}\oi^{-4.7}$. The reason why there are only odd multiples of the base frequency is related to the $\varphi\to -\varphi$ symmetry in the potential \cite{Salmi:2012ta}. (For the regularized logarithmic potential, the leading nonlinear term in the equation of motion starts at $\varphi^3$ due to the $Z_2$ symmetry, which does not support even multiples of the base frequency;  for a potential without the $Z_2$ symmetry, the leading nonlinear term in the equation of motion starts with $\varphi^2$, which supports all multiples of the base frequency.) The two plots in the right column show the evolution of the field and its Fourier spectrum at a point in the outer region of the CSQ. At this point, both the real and imaginary part of the field oscillate in a more complicated way, and the peaks of the spectrum become broader. Also, the dominant peak has two sub-peaks (see the inset of the bottom right plot in Figure \ref{fig:time_fourier_of_CSQs}), which arises because the real and imaginary part of the field have slightly different dominant frequencies, and the difference between the frequencies of the two sub-peaks is approximately the charge-swapping frequency of the CSQ. To see this, note that the field oscillation can be modeled to leading order by $\cphi_r \approx A_r(\mathbf{x}) \cos\(\oi_rt\),~\cphi_i \approx A_i(\mathbf{x}) \cos(\oi_it+\phi_0)$, where $\cphi_r$ and $\cphi_i$ are respectively the real and imaginary part of the scalar field, with $\oi_r$ and $\oi_i$ their dominant frequencies respectively (see Figure \ref{fig:time_fourier_of_CSQs}).  As $\oi_r$ and $\oi_i$ are quite close to each other, we can approximate the 0-th component of the current with $j_0 \approx A_rA_i (\oi_r+\oi_i) \sin\((\oi_r-\oi_i)t - \phi_0\)$, which means that the difference $\oi_i-\oi_r$ is approximately the charge-swapping frequency. 
 
A significant difference between the spectrum plot in the logarithmic potential and that in the $\varphi^6$ case is that the spectra in the logarithmic case is much noisier. This may be easily understood in terms of the form of the potential. While the $\varphi^6$ case only has two nonlinear terms in the equation of motion, the regularized logarithmic potential leads to an equation of motion with an infinite number of nonlinear terms, and these nonlinear terms are large and alternating in signs so as to reproduce the ultra-soft logarithmic interaction. So compared to the $\varphi^6$ case, the higher order nonlinear terms in the logarithmic case can feed many higher order oscillation modes in the spectra, hence more ``noise'' near the peaks.
	
\begin{figure}[tbp]
		\centering
		\includegraphics[width=.45\linewidth]{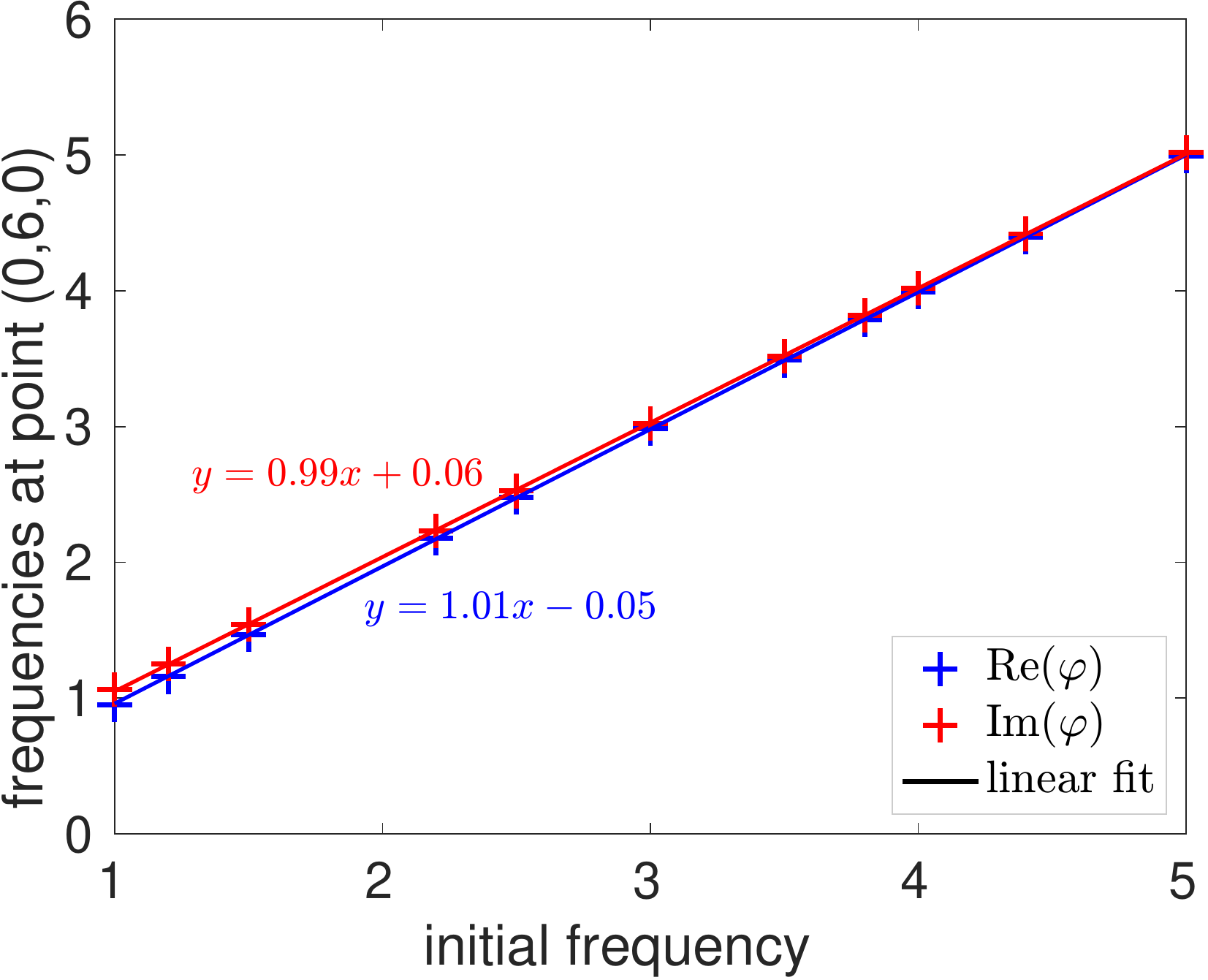}
		\hfill 
		\includegraphics[width=.45\linewidth]{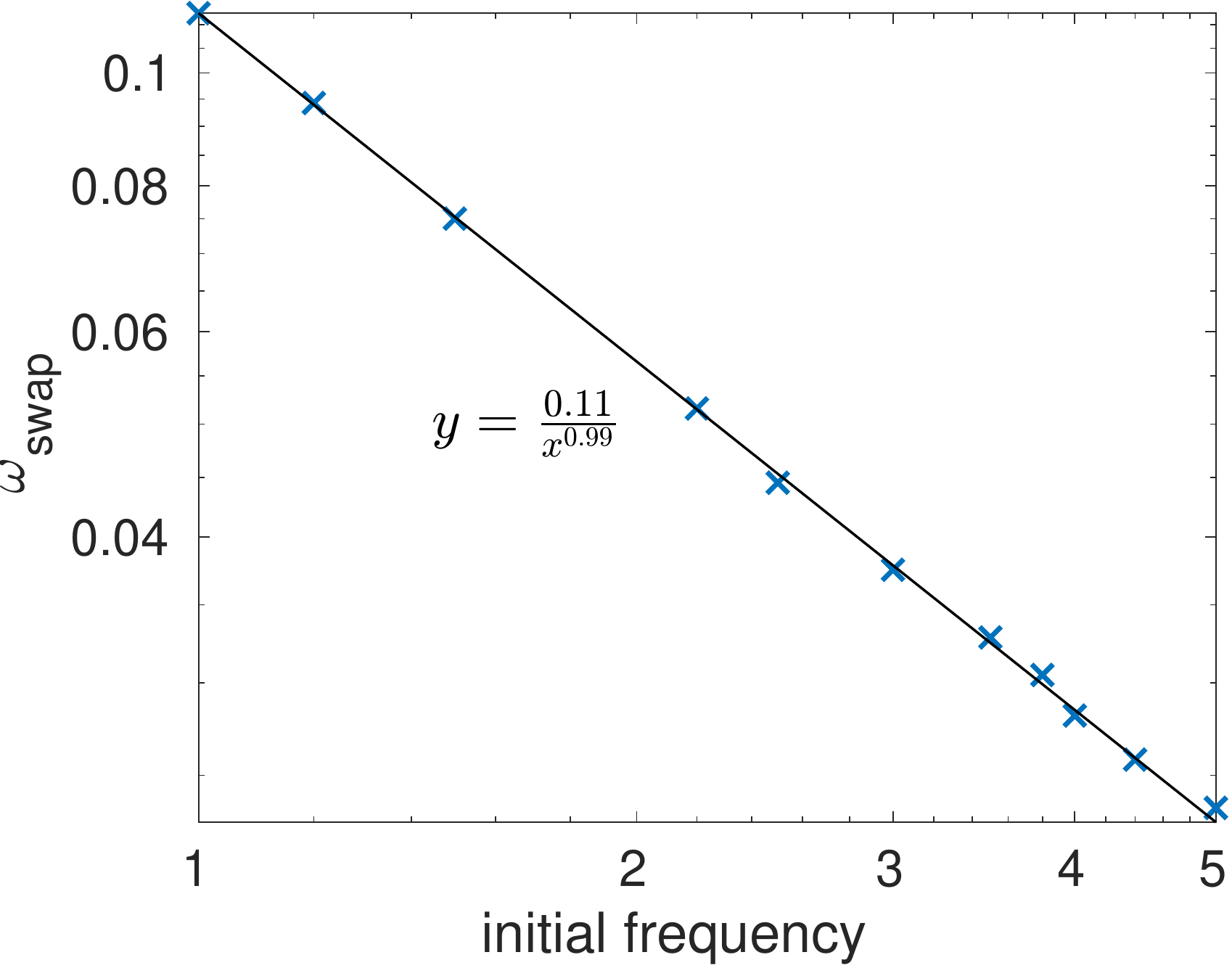}
		\caption{\label{fig:linear_law_of_frequency}
			Dependence of frequencies of the dipole CSQ on the initial frequencies of the constituent Q-balls. The left plot shows the dominant frequencies of the field components at point (0,6,0). Quasi-linear scaling laws are observed for the dominant frequencies of Re($\varphi$) and Im($\varphi$) on the initial frequencies of Q-balls. The right plot shows the dependence of the charge-swapping frequencies on the initial frequencies of the Q-balls. One can check that the difference between the dominant frequencies of Re($\varphi$) and Im($\varphi$) is to a good approximation the charge swapping frequency of the CSQ.
		}
	\end{figure}

Figure \ref{fig:linear_law_of_frequency} illustrates another important difference between the logarithmic case and the $\varphi^6$ case. In the model with the $\varphi^6$ potential, we observed \cite{Xie:2021glp} that a unique dipole CSQ can be obtained with different initial lumps (simple Q-balls with different frequencies/energies or deformed Q-balls or Gaussian lumps) and with different initial separations. Extra energy in the initial configuration will just dissipate away in the initial relaxation process, leaving a CSQ with the same energy and the same charge swapping frequency. As we see in Figure \ref{fig:linear_law_of_frequency}, the situation is rather different in the case of a logarithmic potential, again due to the ultra-soft-ness of the interacting potential. In these plots, we superimpose together a simple Q-ball and a simple anti-Q-ball with different initial frequencies, and examine the resulting CSQ.  From the right plot, we see that the resulting CSQ can have vastly different charge-swapping frequencies, depending on the initial frequencies of the Q-balls. This is an important feature of the logarithmic potential, and probably suggests that the attractors in the logarithmic case form a continuous attractor manifold. 

In the left plot of Figure \ref{fig:linear_law_of_frequency}, we trace the dependence of the dominant frequency of Re($\cphi$) and Im($\cphi$) (at the same edge point as that in Figure \ref{fig:time_fourier_of_CSQs}) on the different initial frequencies of the constituent Q-balls, and find that the dependence is approximately linear. The slopes of the two linear scalings are very close, and the dominant frequency of Re($\cphi$) is always slightly smaller than the initial frequency of the constituent Q-balls, while that of Im($\cphi$) is always slightly larger. Of course, the linear scalings are not exact; otherwise the nonlinear dependence of charge-swapping frequency $\omega_{\rm swap}$ on the initial Q-ball frequencies (the right plot of Figure \ref{fig:linear_law_of_frequency}) can not be reproduced, as the difference between the dominant frequency of Re($\cphi$) and Im($\cphi$) is approximately the charge-swapping frequency.

\subsection{More complex CSQs}
\label{sec:unequal}

In the last subsection, we have mainly focused on the simplest dipole CSQ. In generic circumstances, such as in the scenario of Section \ref{sec:CSQfromAD}, a CSQ may have a much more complex multipolar structure. In this subsection, we will investigate the stability of more complex CSQs. We find that they are still very stable, at least for the simple ones among these complex CSQs. A new feature is that now there are also transit CSQs which are charge swapping for a relatively long time but lose energy faster than normal CSQs.

\begin{figure}[tbp]
    \centering
    \includegraphics[width=0.4\textwidth]{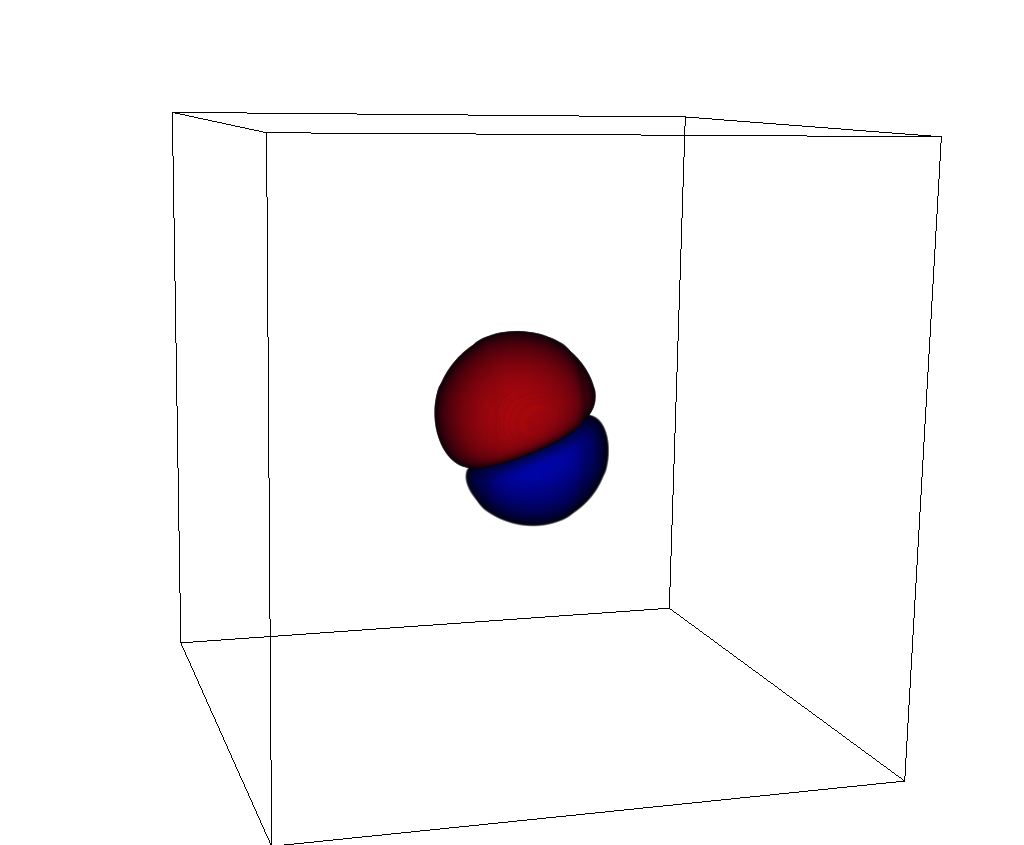}
    \includegraphics[width=0.4\textwidth]{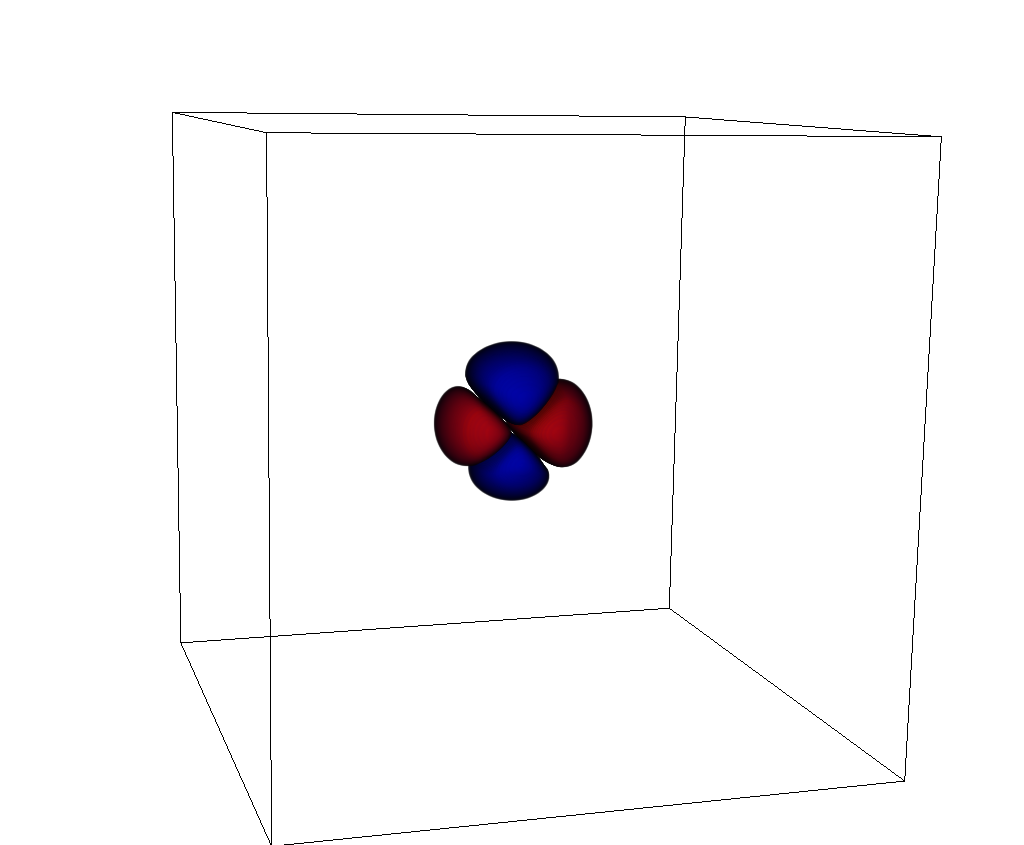}
    \hfill 
    \caption{\label{fig:unequal_3D}
    Charge densities of complex CSQs with unequal charges at $t=1.5\times 10^{5}$. In the left plot, we initially superimpose an anti-Q-ball with $\omega=-2.0$ and a Q-ball with $\omega=2.0$ but with its amplitude deformed by a factor of 1.2, located at $(0,1,0)$ and $(0,-1.0,0)$ respectively. In the right plot, we initially superimpose two of these anti-Q-balls and two of these deformed Q-balls, located symmetrically from the center with a distance of 0.5.
    }
\end{figure}

We may prepare more complex CSQs with unequal positive and negative charges. The simplest case is to superimpose a positive lump and a negative lump with an unequal charge, as in the left plot of Figure \ref{fig:unequal_3D}. In the right plot of Figure \ref{fig:unequal_3D}, on the other hand, we have superimposed four lumps with unequal charges. Both of these cases are still very stable, and we find that their lifetimes are at least $1.5\times 10^{5}/m$, each case taking $ 3.2\times 10^{4} $ CPU hours to simulate. These being shorter than the lifetimes in the previous subsection is due to our limitation of time and computational resources, and it is possible that these unequal complex CSQs are as stable as the dipole or quadrupole CSQs in the previous subsection.

\begin{figure}[tbp]
	\begin{subfigure}{0.22\textwidth}
		\renewcommand{\thesubfigure}{a\arabic{subfigure}}
        \caption{$t=250$}
        \includegraphics[width=\textwidth,trim=160 100 100 150,clip]{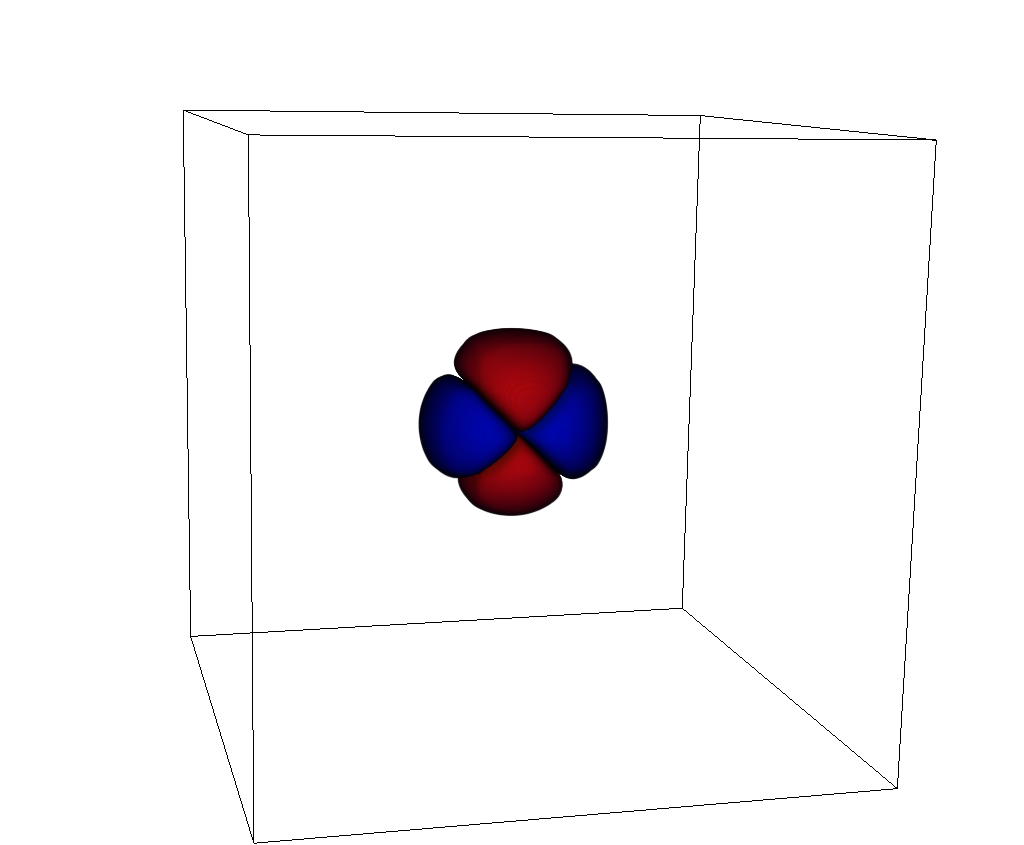}
    \end{subfigure}
    \hfill 
    \begin{subfigure}{0.22\textwidth}
		\renewcommand{\thesubfigure}{a\arabic{subfigure}}
        \caption{$t=2000$}
        \includegraphics[width=\textwidth,trim=160 100 100 150,clip]{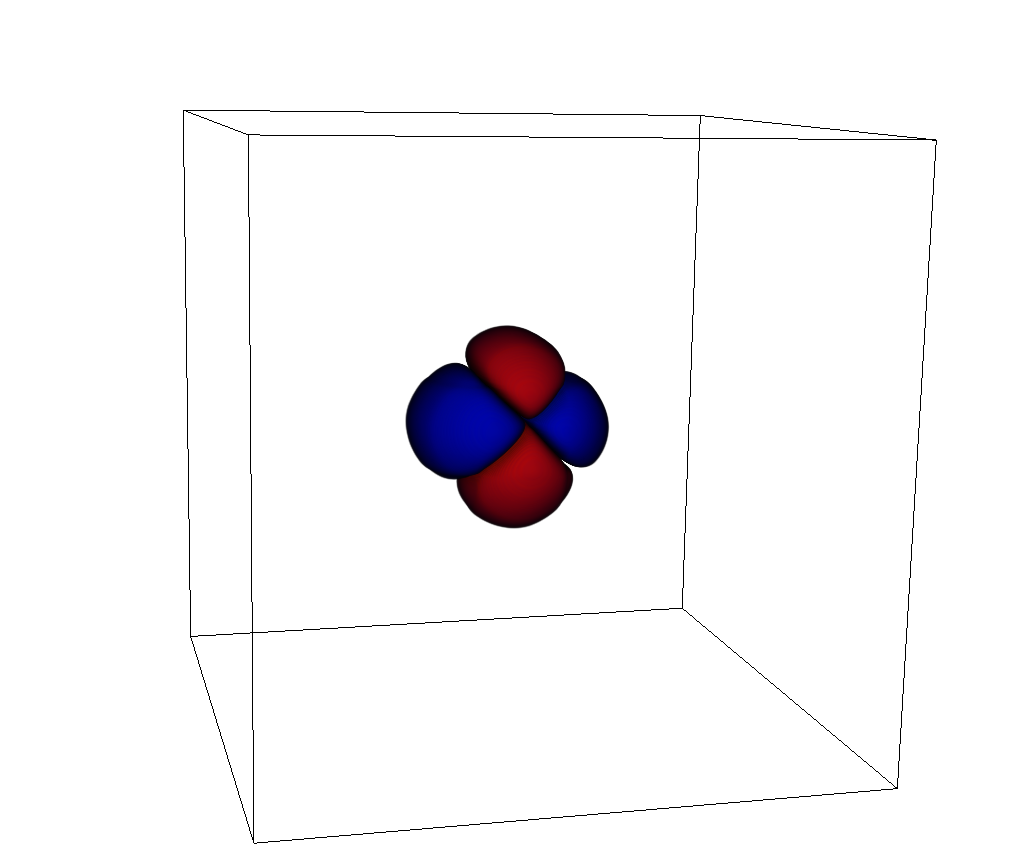}
    \end{subfigure}
    \hfill 
    \begin{subfigure}{0.22\textwidth}
		\renewcommand{\thesubfigure}{a\arabic{subfigure}}
        \caption{$t=4000$}
        \includegraphics[width=\textwidth,trim=160 100 100 150,clip]{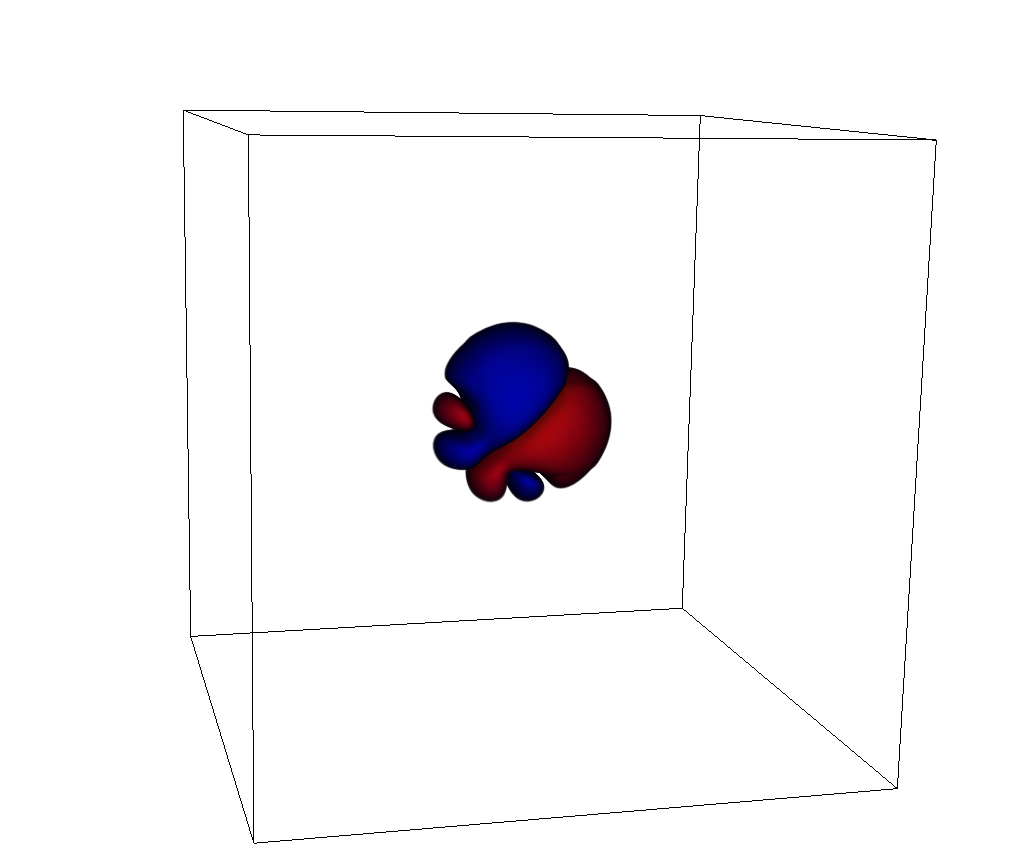}
    \end{subfigure}
    \hfill 
    \begin{subfigure}{0.22\textwidth}
		\renewcommand{\thesubfigure}{a\arabic{subfigure}}
        \caption{$t=7000$}
        \includegraphics[width=\textwidth,trim=160 100 100 150,clip]{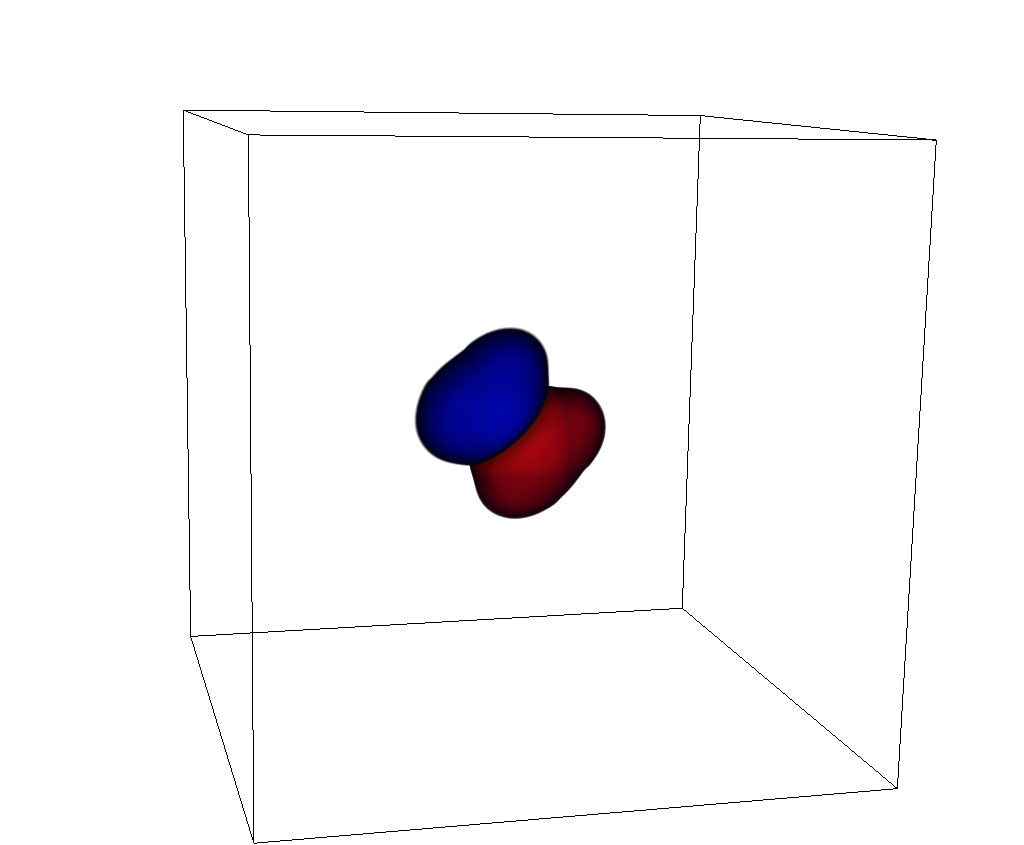}
    \end{subfigure}
    \begin{subfigure}{0.22\textwidth}
		\setcounter{subfigure}{0}
		\renewcommand{\thesubfigure}{b\arabic{subfigure}}
        \caption{$t=25$}
        \includegraphics[width=\textwidth,trim=160 100 100 150,clip]{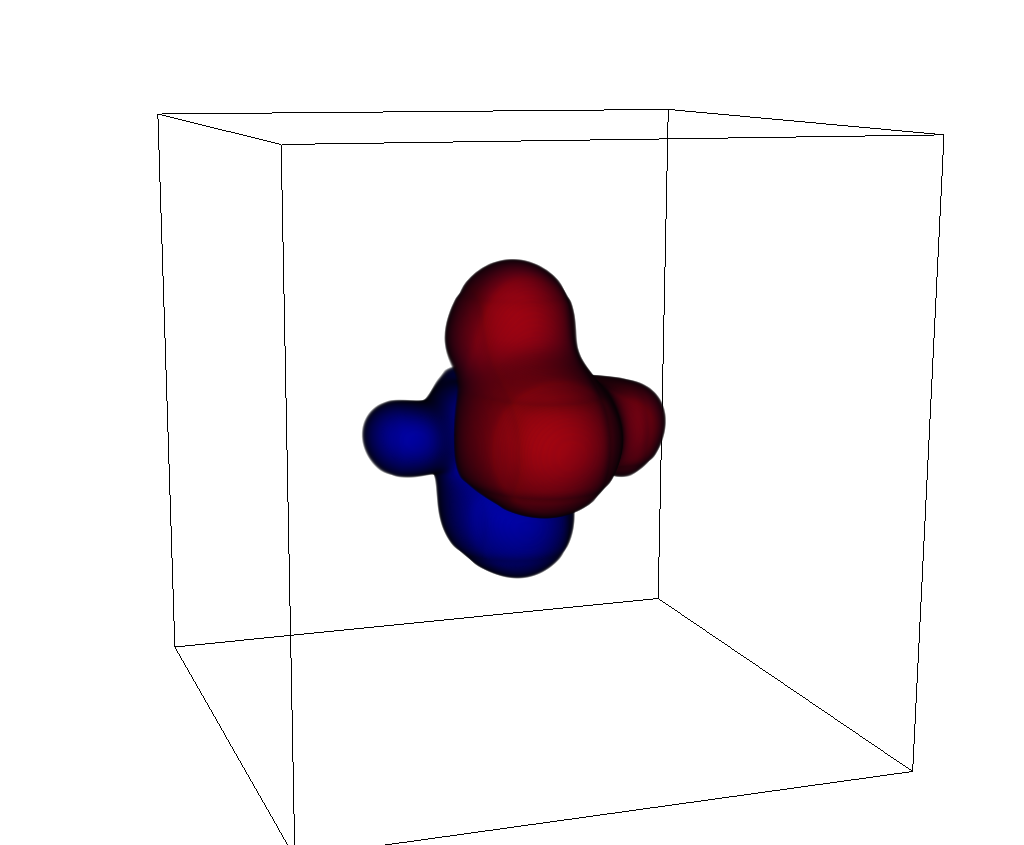}
    \end{subfigure}
    \hfill 
    \begin{subfigure}{0.22\textwidth}
		\renewcommand{\thesubfigure}{b\arabic{subfigure}}
        \caption{$t=1500$}
        \includegraphics[width=\textwidth,trim=160 100 100 150,clip]{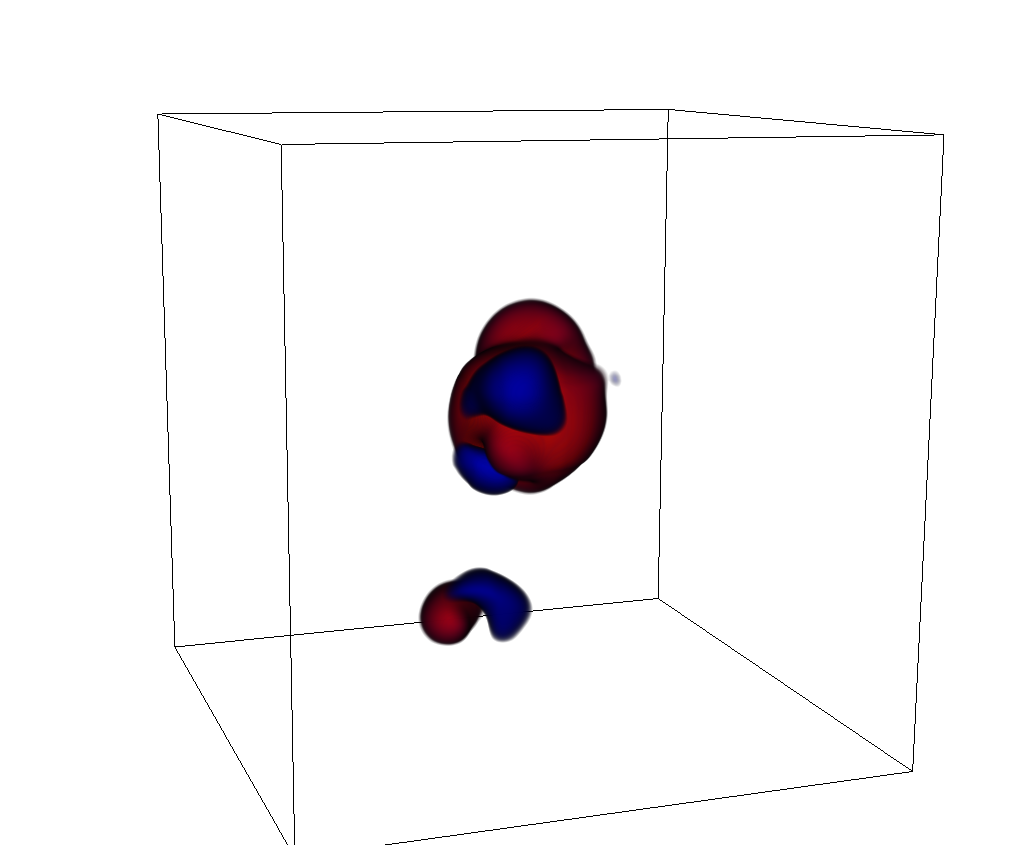}
    \end{subfigure}
    \hfill 
    \begin{subfigure}{0.22\textwidth}
		\renewcommand{\thesubfigure}{b\arabic{subfigure}}
        \caption{$t=3000$}
        \includegraphics[width=\textwidth,trim=160 100 100 150,clip]{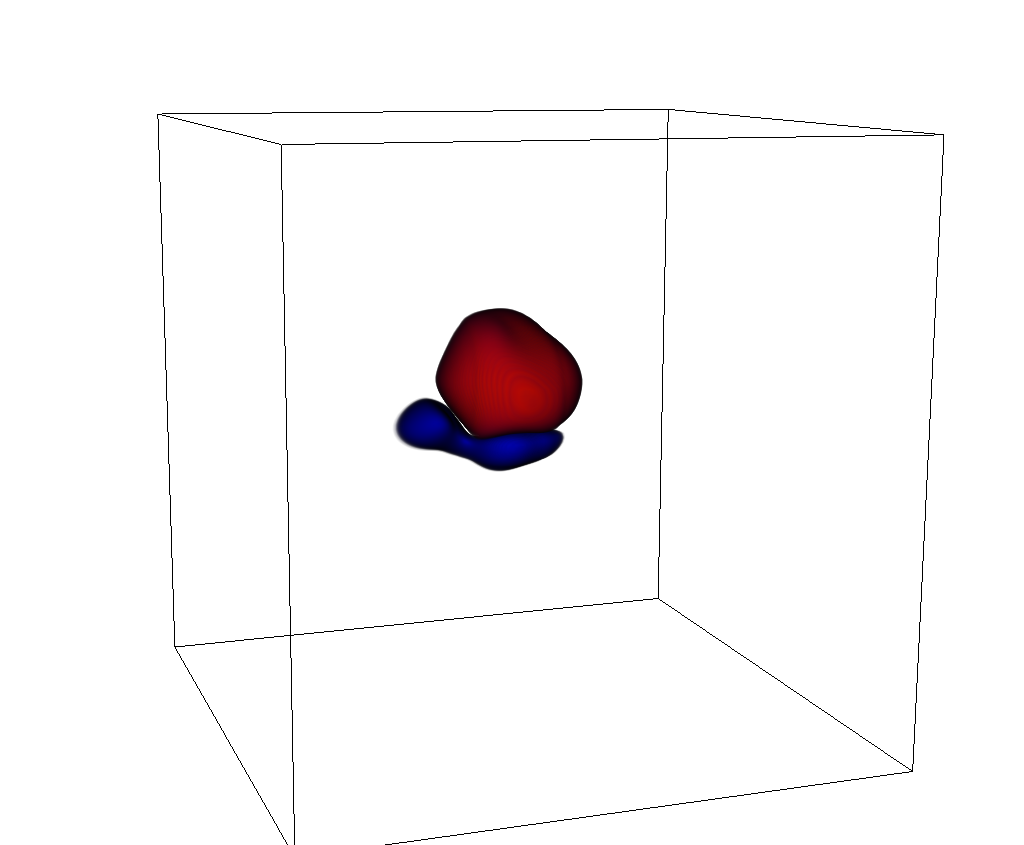}
    \end{subfigure}
    \hfill 
    \begin{subfigure}{0.22\textwidth}
		\renewcommand{\thesubfigure}{b\arabic{subfigure}}
        \caption{$t=5000$}
        \includegraphics[width=\textwidth,trim=160 100 100 150,clip]{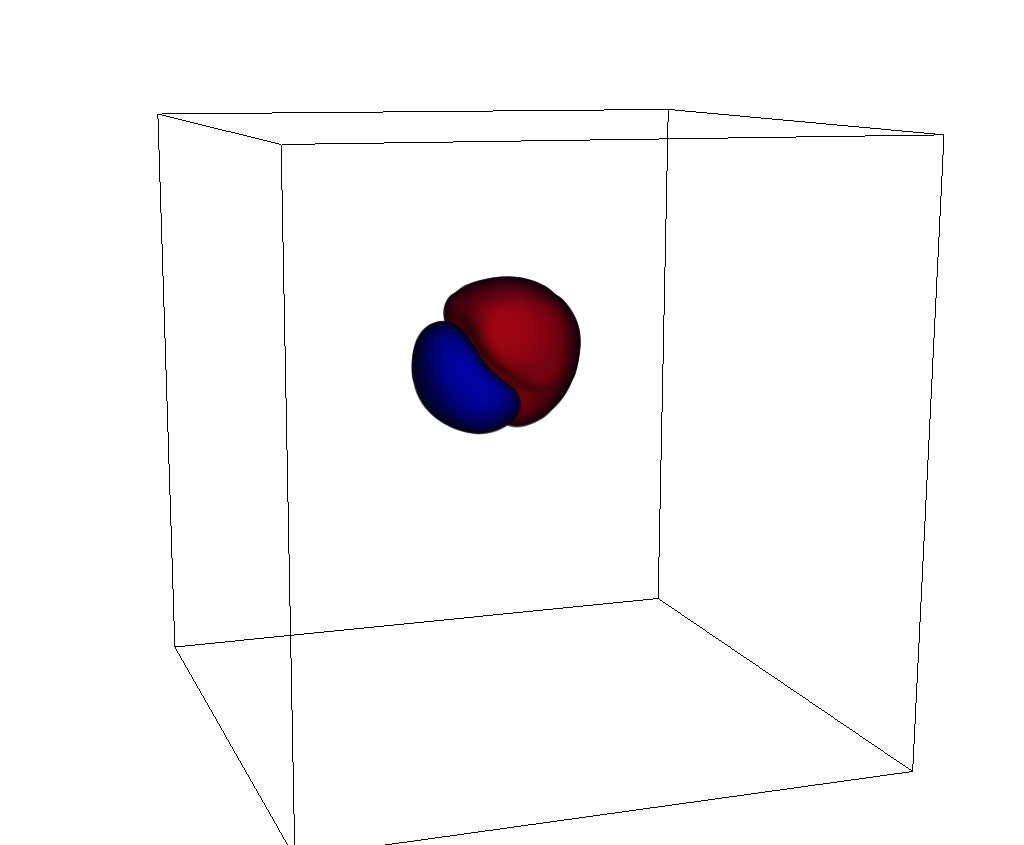}
    \end{subfigure}
    \caption{\label{fig:evolution_sequence_of_complex_CSQs}
        Two evolution sequences of ``transit CSQs'' in 3+1D. These transit states continuously emit radiation and evolve to an unequal CSQ with only one positive and one negative charge lump. In the first row, 4 unequal charge lumps are initially placed at a distance of $2.0$ from the center, with other parameters the same as those in the right plot of Figure \ref{fig:unequal_3D}. In the second row, the initial configuration is prepared with 25 charge lumps, rather randomly displaced.
    }
\end{figure}

On the other hand, if we prepare the initial configuration in the right plot of Figure \ref{fig:unequal_3D} with slightly larger distances between the constituent lumps,  we will end up with transit CSQs. Charges are still swapped within these transit states, but they continuously emit larger amounts of radiation and their total energies do not plateau as normal CSQs would do. With time, they evolve to an unequal CSQ with just one lump of positive charge and one lump of negative charge, just like the left plot of Figure \ref{fig:unequal_3D}. In fact, a transit CSQ decaying to an unequal CSQ with two lumps is usually what we encounter when we want to prepare more complex unequal CSQs via superposition of charge lumps. See Figure \ref{fig:evolution_sequence_of_complex_CSQs}. While the superposition method is useful to obtain CSQs with fixed multipoles (dipole, quadruple, etc.), it does not seem to be reliable to construct more complex CSQs. On the other hand, as we see in Section \ref{sec:CSQfromAD}, complex CSQs can actually arise naturally in a preheating-like setup, but in that case, we have little control over what kinds of complex CSQs are produced. We leave the search for alternative methods to construct complex CSQs for future work.

\subsection{Attractor basin of CSQ formation}
\label{sec:attractorBasin}
	
As the CSQs are quasi-stable configurations, they must have an attractor basin of formation. In this subsection, we investigate how easily logarithmic CSQs can form from colliding simple Q-balls. Since many salient dynamical properties of logarithmic CSQs in 3D are very similar to the ones in 2D, we will scan the parameter space in 2D for simplicity.

\begin{figure}[tbp]
		\centering
		\includegraphics[width=.6\textwidth]{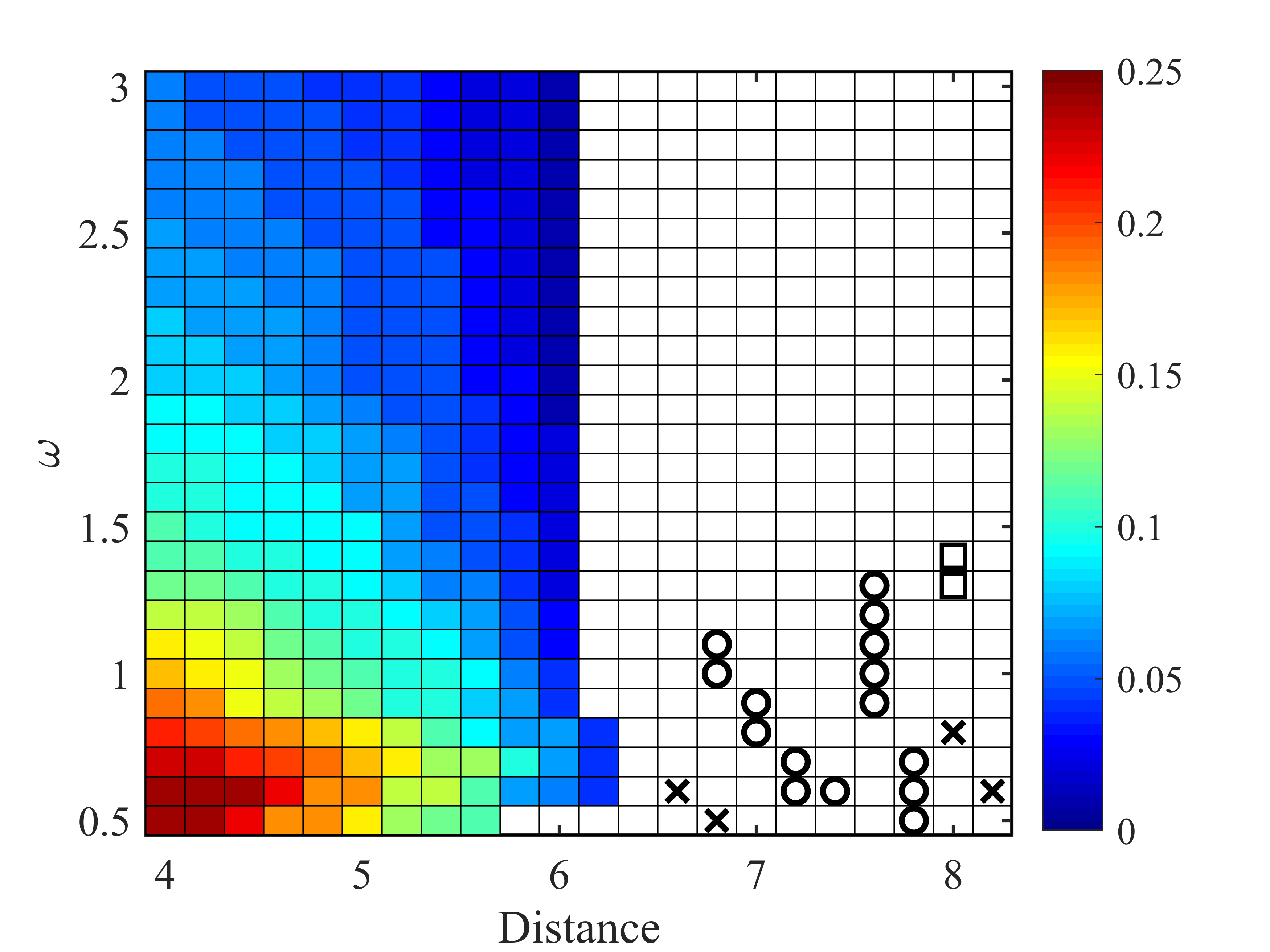}
		\caption{\label{fig:formation_of_CSQs_w-d_sample}
	           Parameter space where a CSQ can form for a head-on collision between a Q-ball and an anti-Q-ball. The horizontal axis is the distance between the center of the initial Q-ball and anti-Q-ball, the vertical axis is the absolute value of the frequency of the initial Q-ball and anti-Q-ball, and the color bar denotes the maximum initial velocities allowed for a CSQ to form (251 bins for color values). The black symbols (crosses, circles and boxes) on the right are the special cases where a Q-ball and an anti-Q-ball with zero initial velocities collide to form two CSQs moving in opposite directions. The crosses, circles and boxes correspond to the cases where there are respectively 5, 3 and 2 charge swaps in the central lump before the two CSQs are formed and move away.
		}
	\end{figure}

First, we consider the case where a Q-ball and an anti-Q-ball that have opposite frequencies $\oi$ and $-\oi$ and no relative phase are initially placed at a distance apart with a relative velocity, and then they collide head-on. (To get a Q-ball solution with a velocity, we can Lorentz boost the solution in Eq.~(\ref{varphiini}).) In Figure \ref{fig:formation_of_CSQs_w-d_sample} we plot the parameter space for this case where a CSQ can form. The simulation box is in a frame where the Q-ball and the anti-Q-ball have the same speed and opposite direction. The maximum initial Q-ball velocity (with respect to the simulation box) that can form a CSQ for the corresponding initial distance and frequency is plotted with different colors, and the white region is where a CSQ can not form for any velocity. We can see that whether a CSQ can form mainly depends on the initial distance between the center of the two Q-balls. The two Q-balls should be close so that their nonlinear cores have some overlap for the CSQ to form. Also, when the frequency $\oi$ or the distance is smaller, a large initial velocity is allowed before they can not form a CSQ. Notice that a Q-ball and an anti-Q-ball with appropriate phases attract each other when placed far apart, so a large initial distance is to some extent similar to a large initial velocity. The black data points (crosses, circles and boxes) on the right of Figure \ref{fig:formation_of_CSQs_w-d_sample} are the special cases where a Q-ball and an anti-Q-ball, both of which have zero velocity initially, collide to form two CSQs moving away from each other. (The colored region on the left is when the Q-ball and anti-Q-ball collision results in only one CSQ.) In these cases, before two CSQs are formed and move away in opposite directions, the positive and negative charges can swap for a number of times. The crosses, circles and boxes correspond to the cases where the number of swaps is 5, 3 and 2 respectively. 
Just outside the boundary of the colored region, positive and negative charges can also swap for a number of times, but during these swaps all the energies in the lump are quickly radiated away. 
	
	\begin{figure}[tbp]
		\centering
		\includegraphics[width=.95\textwidth]{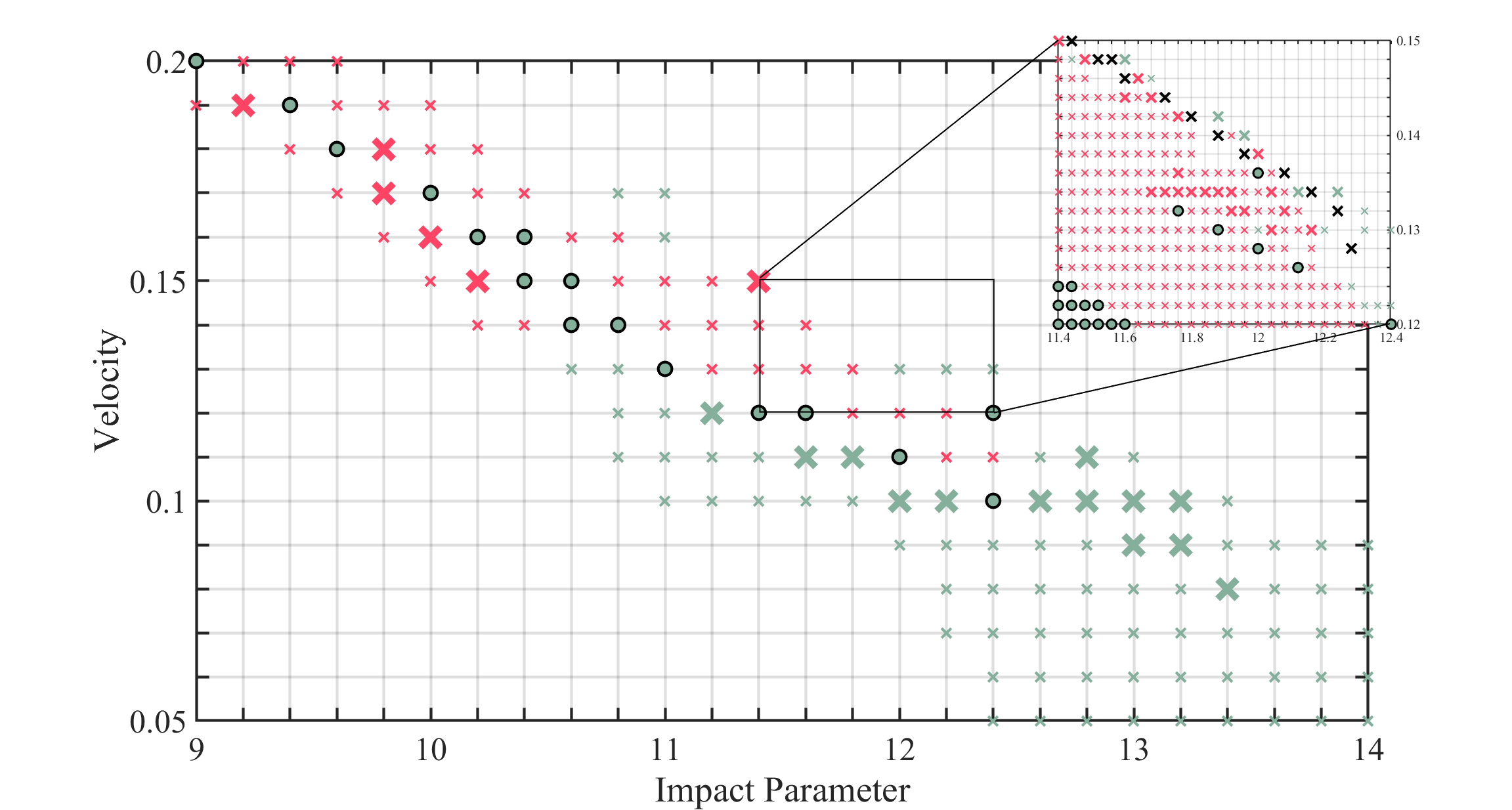}
		\caption{\label{fig:formation_of_CSQs_v-d_sample}
			Parameter space where CSQs can form for a non-head-on collision between a Q-ball and an anti-Q-ball. The horizontal axis is the impact parameter between the centers of the Q-ball and anti-Q-ball, and the vertical axis is the initial velocities of the two balls. The scattering results are shown in Figure \ref{fig:consequences_of_off-axis_collision}: (a) two balls passing through, (b) two CSQs (green crosses), (c) one CSQs and two simple Q-balls (red crosses), (d) three CSQs (green dots), (e) one CSQ (black crosses). Smaller crosses mean that the resulting CSQs have noticeably less charges. The frequencies of the Q-ball and the anti-Q-ball are $1.0$ and $-1.0$ respectively, and the initial positions of the two balls are at $x=\pm\,\mathrm{ImpactParameter}/2$ and $y=\pm 10$ respectively.
		}
	\end{figure}

	\begin{figure}[tbp]
		\centering
		\begin{subfigure}{0.32\textwidth}
			\caption{}
			\includegraphics[width=\textwidth]{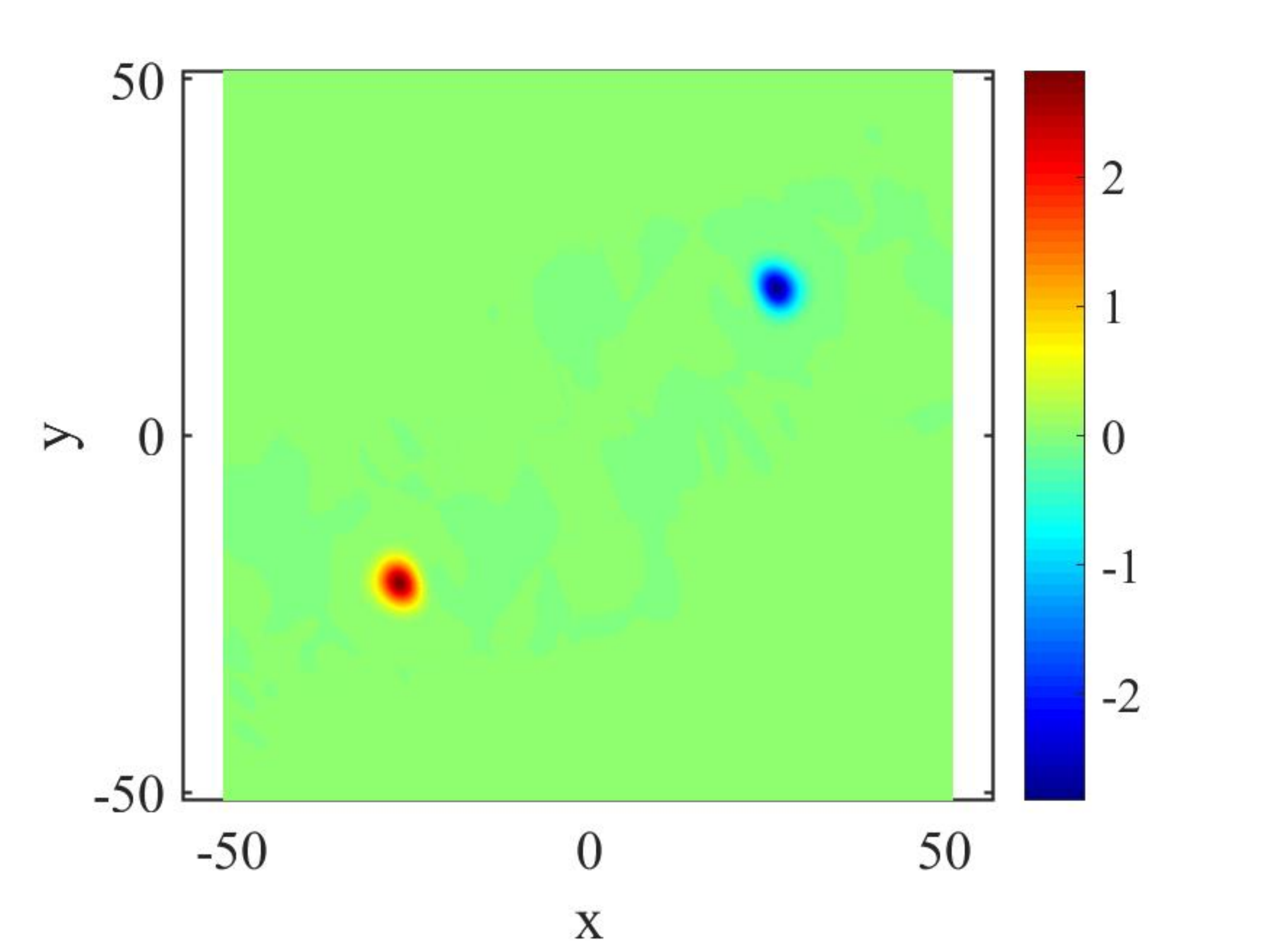}
		\end{subfigure}
		\hfill
		\begin{subfigure}{0.32\textwidth}
			\caption{}
			\includegraphics[width=\textwidth]{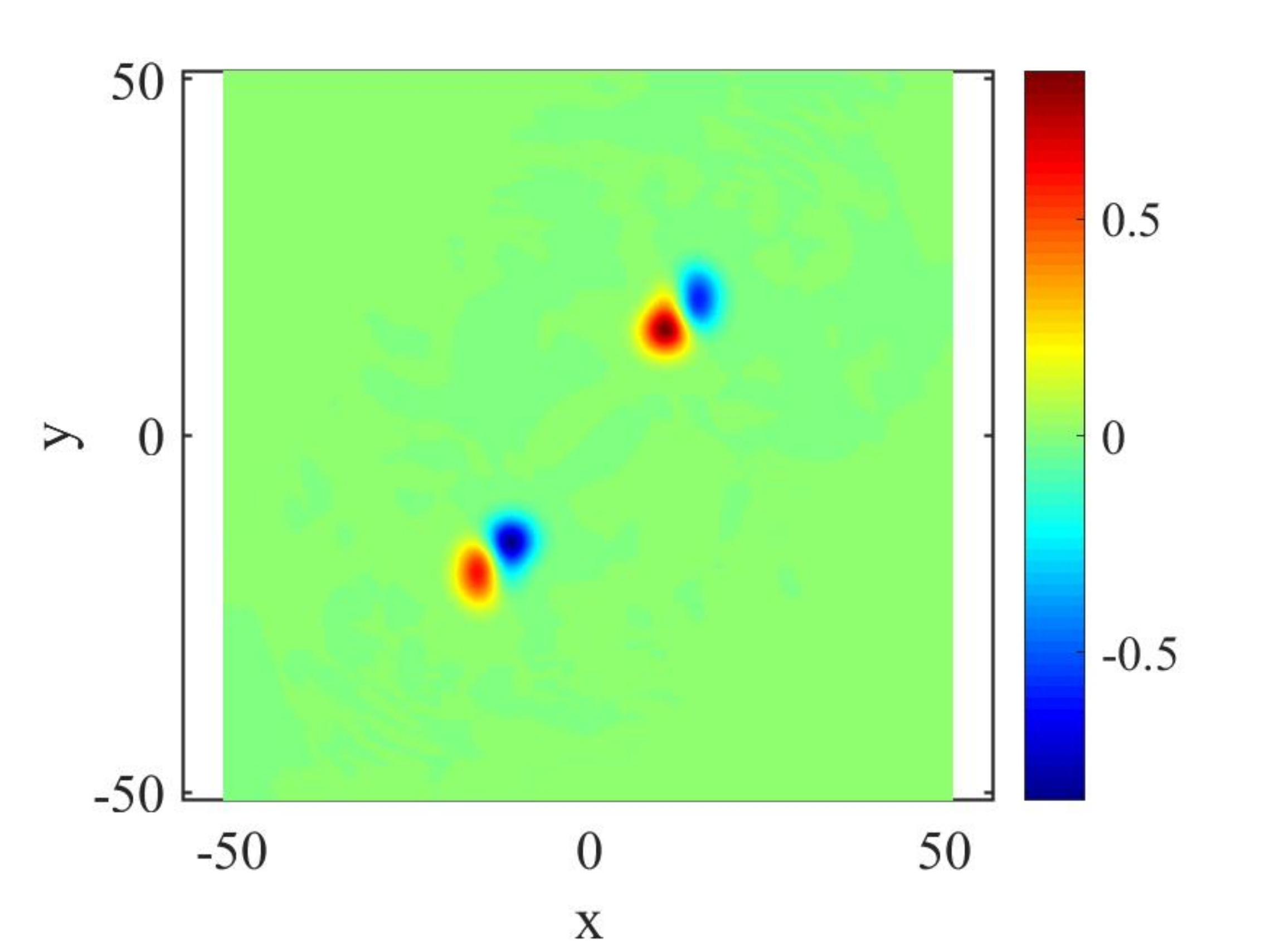}
		\end{subfigure}
		\hfill
		\begin{subfigure}{0.32\textwidth}
			\caption{}
			\includegraphics[width=\textwidth]{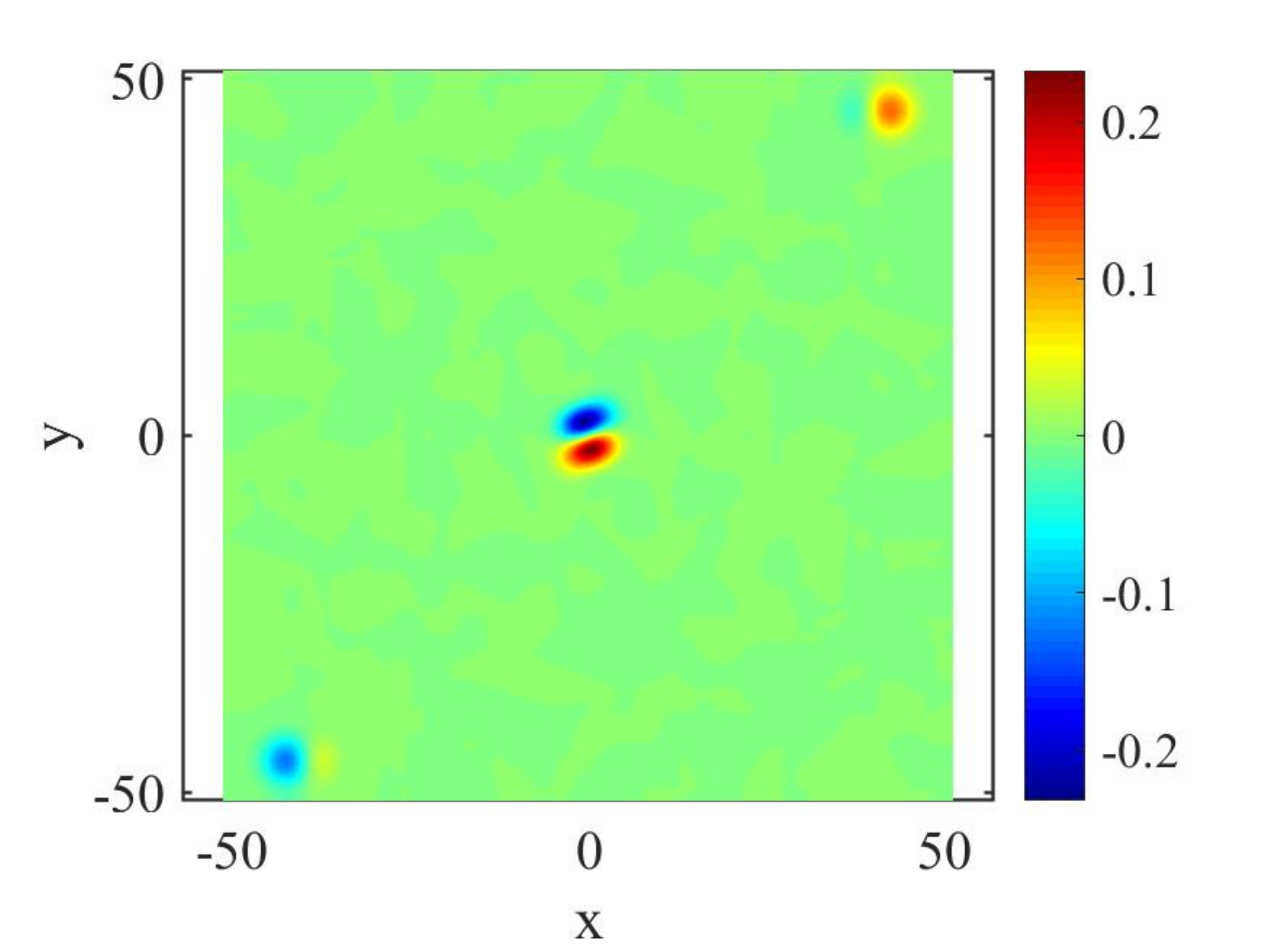}
		\end{subfigure}
		\hfill
		\begin{subfigure}{0.32\textwidth}
			\caption{}
			\includegraphics[width=\textwidth]{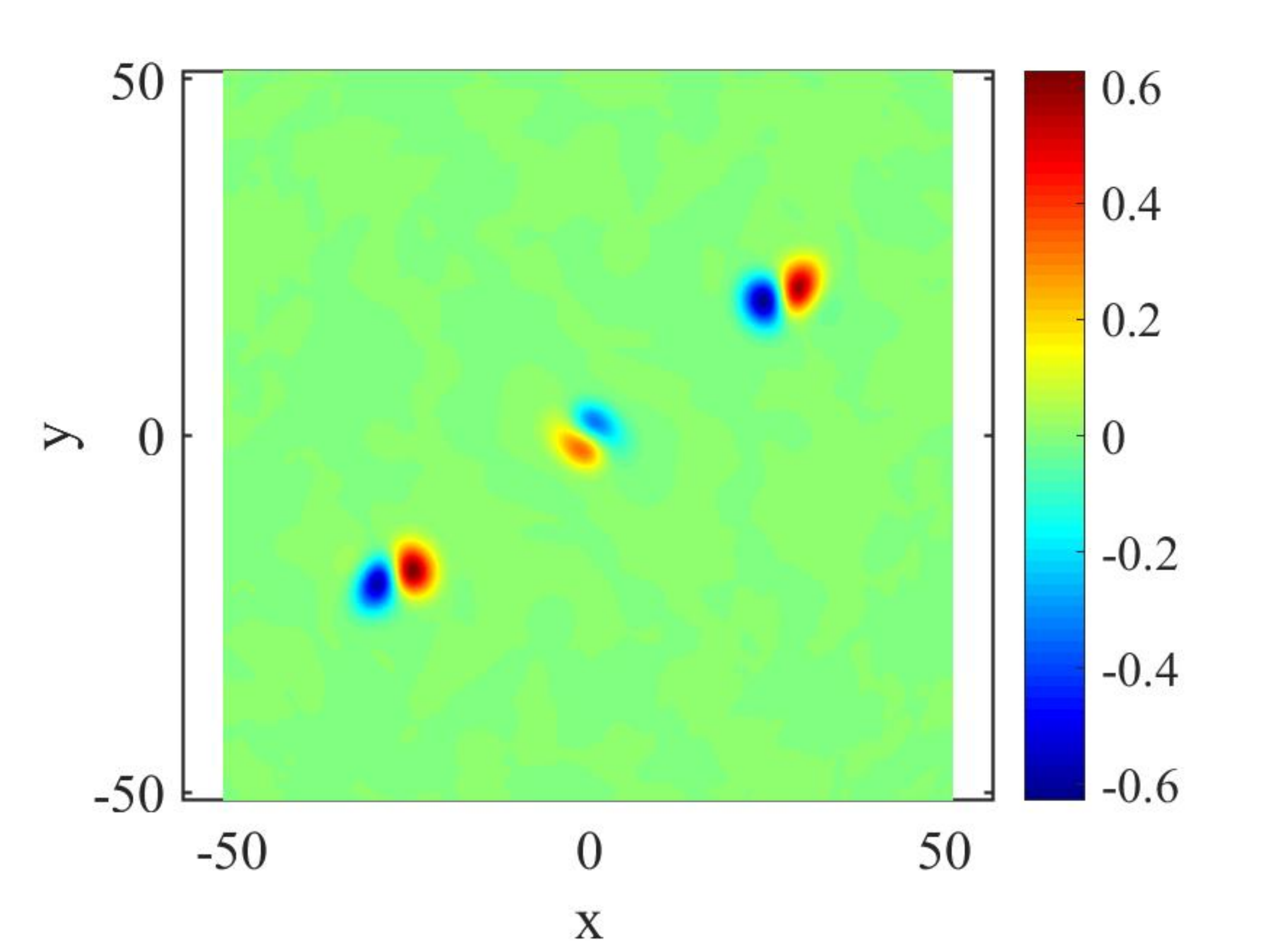}
		\end{subfigure}
		~~~~
		\begin{subfigure}{0.32\textwidth}
			\caption{}
			\includegraphics[width=\textwidth]{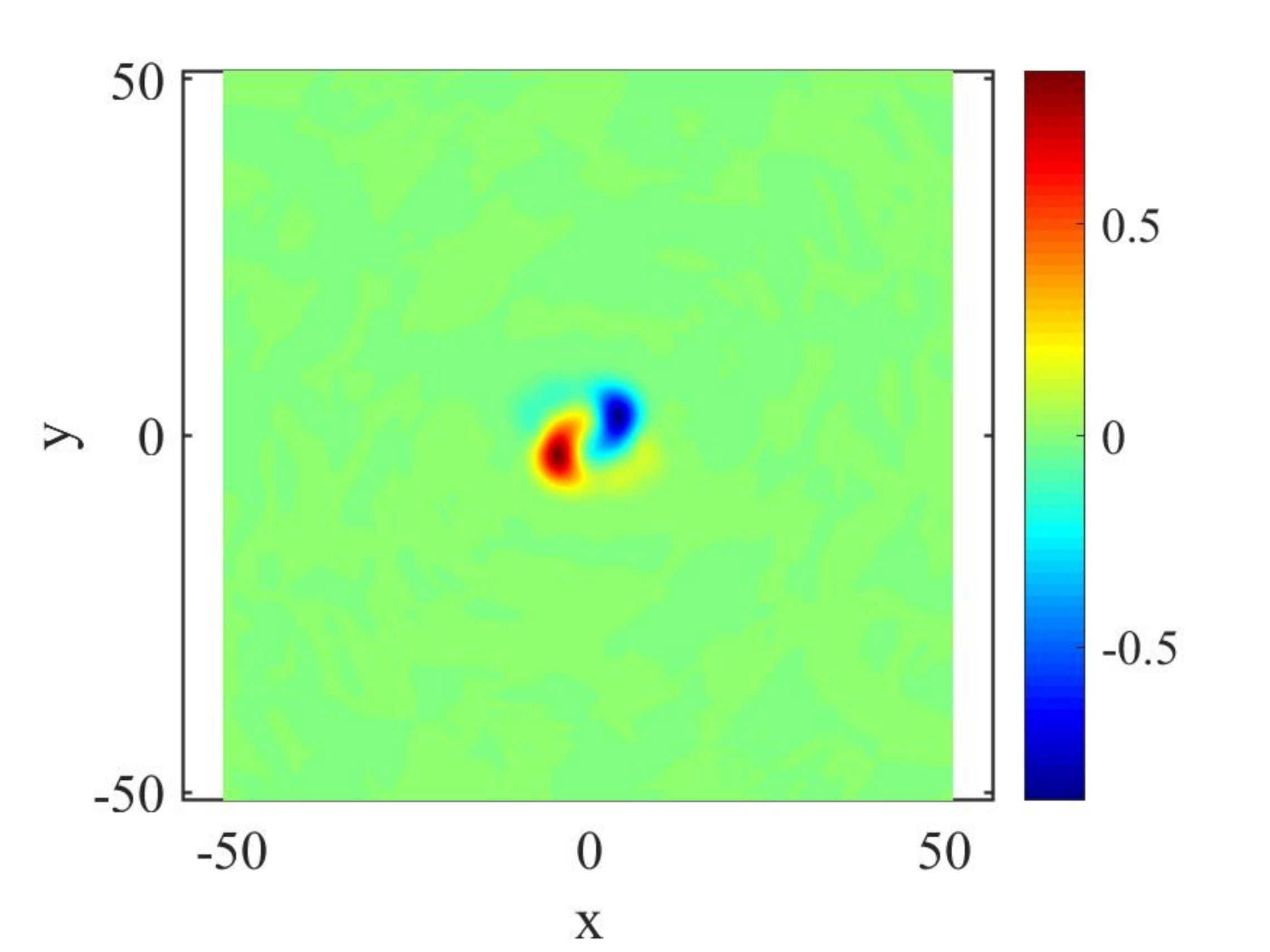}
		\end{subfigure}
		\hfill
		\caption{\label{fig:consequences_of_off-axis_collision}
			Different end-states for the non-head-on collisions used to collate the data in Figure \ref{fig:formation_of_CSQs_v-d_sample}, the color represents the charge density.
		}
	\end{figure}
	
For Q-balls that are initially far away or collide with large initial velocities, they will pass through each other in head-on collisions. However, it may be expected that non-head-on collisions are more probable in realistic situations, in which case CSQs can still form for large initial separations or velocities. In Figure \ref{fig:formation_of_CSQs_v-d_sample} we chart the parameter space of CSQ formation for different initial velocities and impact parameters in non-head-on collisions. We see that in this velocity vs impact parameter plot it is the diagonal strip that supports CSQ formation, \ie when the initial velocity is smaller, the impact parameter has to be larger, and vice versa. Not surprisingly, the results have many different end-states for the non-head-on collisions. When the initial conditions are unfavorable, the two balls just pass by each other. In the diagonal strip of the plot, the end state of the collision can be two or three CSQs, or one CSQ and two simple Q-balls, Fig. \ref{fig:consequences_of_off-axis_collision}. As shown in the inset of Figure \ref{fig:formation_of_CSQs_v-d_sample}, we find that it is actually not rare to see three CSQs or one single CSQ as the end state. In the latter case, the positive and negative charges in the resulting single CSQ rotate around each other, much like the Tai Chi diagram.

\section{Summary}
	\label{sec:Summary}
	
CSQs are composite Q-ball solutions that are localized and long-lived, within which positive and negative charges swap with a well-defined frequency that is smaller than the oscillation frequency of the field. They exist in theories where the simple spherically symmetric Q-balls exist, and yet their fascinating properties are just starting to be uncovered. In this paper, we have studied the properties of CSQs in a U(1) scalar model with a logarithmic potential, which has an interesting bearing in the early universe. Indeed, we have shown that complex CSQs can be copiously generated in the Affleck-Dine fragmentation process via parametric resonance from small random initial perturbations. This is of course consistent with the fact that CSQs are attractor meta-stable solutions.  We have then investigated logarithmic CSQs prepared in more controlled ways.

The logarithmic potential is special because it is ultra-soft, which means that the logarithmic CSQs are extremely stable. We would only expect them to be quasi-stable, as the simple Q-ball solutions have lower energies for the same charge, but in our parallelized long-term simulations we have not seen the decay of properly prepared CSQs, either in 3D or in 2D. This is also irrespective of whether the charge lumps in the CSQ are equal or not. To determine the lifetimes of CSQs is challenging with long-time lattice simulations because the lattice has a finite size. To minimize unwanted perturbations in the simulation box, we impose the 2nd order Higdon absorbing boundary conditions. We have also charted the attractor basin to form the CSQs from two colliding Q-balls, either head-on or non-head-on, for three parameters: the frequency of the Q-balls, the initial separation and the colliding velocity.

Let us summarize the main differences between the logarithmic CSQs and the sextic CSQs \cite{Xie:2021glp}. First of all, the lifetime of a sextic CSQ is typically shorter than that of the logarithmic CSQ, which allowed us to follow the whole evolution history of a sextic CSQ. In the first relaxation stage of a sextic CSQ, a large amount of energy sheds away before entering the CSQ stage when the characteristic charge-swapping feature appears, while for the logarithmic case well-defined charge swapping appears from the get-go. Also, there can be many charge-swapping frequencies for the logarithmic CSQs depending on the initial conditions, while the sextic CSQs only have a unique charge-swapping frequency for various different initial conditions. All of these are of course ultimately because of the softness of the logarithmic potential. It is expected that these differences persist if one is to compare the logarithmic CSQs and other CSQs with a typical polynomial potential.

\acknowledgments
	
We would like to thank Xiao-Xiao Kou and Yi-Jie Wang for helpful discussions. SYZ acknowledges support from the starting grants from University of Science and Technology of China under grant No.~KY2030000089 and GG2030040375, and is also supported by National Natural Science Foundation of China under grant No.~11947301, 12075233 and 12047502, and supported by the Fundamental Research Funds for the Central Universities under grant No.~WK2030000036.  The work of PMS was funded by STFC Consolidated Grant Number ST/T000732/1. We would like to acknowledge the use of the public software VAPOR and VisIt for volume rendering.
\\

	\bibliographystyle{JHEP}
	\bibliography{refs}

\end{document}